\documentclass[pdflatex,sn-mathphys-num]{sn-jnl} 

\usepackage{stmaryrd}
\usepackage{subcaption}
\usepackage{floatrow}

\usepackage{graphicx}%
\usepackage{multirow}%
\usepackage{amsmath,amssymb,amsfonts}%
\usepackage{amsthm}%
\usepackage{mathrsfs}%
\usepackage[title]{appendix}%
\usepackage{xcolor}%
\usepackage{textcomp}%
\usepackage{manyfoot}%
\usepackage{booktabs}%
\usepackage{listings}%
\usepackage{array}
\usepackage{placeins}
\usepackage[lined,boxed,ruled,commentsnumbered]{algorithm2e}

\newtheorem{theorem}{Theorem}
\newtheorem{lemma}{Corollary}

\newcommand{\pder}[2]{\ensuremath{\frac{\partial #1}{\partial #2}}}

\newcommand{\PreserveBackslash}[1]{\let\temp=\\#1\let\\=\temp}
\newcolumntype{C}[1]{>{\PreserveBackslash\centering}p{#1}}
\newcolumntype{R}[1]{>{\PreserveBackslash\raggedleft}p{#1}}
\newcolumntype{L}[1]{>{\PreserveBackslash\raggedright}p{#1}}

\newcommand{\change}[1]{\textcolor{black}{#1}} 

\newcommand{\changee}[1]{\textcolor{black}{#1}} 

\begin{document}

\title[Adjoint-based optimization of the Rayleigh-Bénard instability with melting boundary]{Adjoint-based optimization of the Rayleigh-Bénard instability with melting boundary}


\author*[1,2]{\fnm{Tomas} \sur{Fullana}}
\email{tomas.fullana@gmail.com}

\author[1]{\fnm{Alejandro} \sur{Quir{\'o}s Rodr{\'i}guez}}

\author[3]{\fnm{Vincent} \sur{Le Chenadec}}

\author*[1,4]{\fnm{Taraneh} \sur{Sayadi}}
\email{taraneh.sayadi@lecnam.net}

\affil[1]{\orgdiv{Institut Jean le Rond d'Alembert}, \orgname{Sorbonne Universit{\'e}, CNRS}, \orgaddress{\city{Paris}, \postcode{F-75005}, \country{France}}}

\affil[2]{\orgdiv{Laboratory of Fluid Mechanics and Instabilities}, \orgname{EPFL}, \orgaddress{\city{Laussane}, \postcode{CH-1015}, \country{Switzerland}}}

\affil[3]{\orgdiv{MSME}, \orgname{Universit\'{e} Gustave Eiffel, UPEC, CNRS}, \orgaddress{\city{Marne-la-Vall{\'e}e}, \postcode{F-77454}, \country{France}}}

\affil[4]{\orgdiv{Mathematical and Numerical Modelling Laboratory}, \orgname{Conservatoire National des Arts et M{\'e}tiers}, \orgaddress{\city{Paris}, \postcode{F-75003}, \country{France}}}


\abstract{In this work, we propose an adjoint-based optimization procedure to control the onset of the Rayleigh-Bénard instability with a melting front. A novel cut cell method is used to solve the Navier-Stokes equations in the Boussinesq approximation and the convection-diffusion equation in the fluid layer, as well as the heat equation in the solid phase. To track the interface we use the level set method where its evolution is simply governed by an advection equation. \changee{An incomplete continuous adjoint problem is then derived by treating the velocity field obtained from the forward problem as a known variable in the adjoint convection-diffusion equation, thereby avoiding the need to solve a Navier–Stokes adjoint in the fluid phase. To the best of our knowledge, this provides the first adjoint-based optimization framework for Rayleigh–Bénard instability with a melting boundary. Two optimization problems, together with a comparison against a derivative-free particle–swarm method, demonstrate that the proposed incomplete adjoint yields gradients accurate enough to control the front shape while reducing the number of expensive function evaluations by about an order of magnitude.}}


\maketitle

\section{Introduction}\label{sec:intro}

Stefan problem~\cite{sarler_stefans_1995} models transport and transfer phenomena, in particular solid-liquid phase change in evaporating or chemically reacting flows. Such phenomena govern the interface motion in many engineering related problems such as dendritic solidification~\cite{osher_fronts_1988,Juric1996}, phase transformation in metallic alloys~\cite{Segal1998}, and solid fuel combustion~\cite{Hassan2021}.  
At the interface, the Stefan condition arises from the interaction of liquid and solid phases (both considered incompressible), resulting in a moving liquid-solid interface (freezing or melting front). The speed of the front is directly related to the jump in the conductive heat flux across the interface.  In one dimension, this problem has been studied in depth~\cite{javierre_comparison_2006,Rose1993,Brattkus1992}; in higher dimensions, however, due to the unstable pattern formation~\cite{langer_instabilities_1980,mullins_stability_1964,woods_melting_1992} specific numerical methods, such as the level set method~\cite{Limare2022, osher_fronts_1988,gibou_level_2003,chen_numerical_2009,jenkins_level_2015, bayat_sharp_2022, fullana2022, silva_topology_2022, Fullana2023, QuirosRodriguez2024, castanar_topology_2024}, have been used to predict the motion and shape of the interface. In crystal growth, for example, under-cooling triggers an instability mechanism, causing the solid phase of the material to grow into the liquid phase in a finger-like or dendritic fashion, resulting in complex interfacial shapes and possible topography changes. Moreover, while convection with a fixed topography is usually studied in the classical Rayleigh-Bénard setup~\cite{bodenschatz_recent_2000}, many natural phenomena such as erosion or melting involve a coupling between the flow and the moving boundary~\cite{Yang_Howland_Liu_Verzicco_Lohse_2023, perissutti2024morphodynamicsmeltingiceturbulent}.

The shape of the resulting interface may strongly affect the outcome and time-frame of the production processes in many industrial applications involving phase change~\cite{picelli_topology_2020, siqueira_topology_2024}. As a result, it is desirable to extract efficient control strategies to manipulate the motion of the interface, for instance, by tracking a prescribed trajectory. Two major types of optimization methods in use today are (i) gradient-based, and (ii) derivative free methods. While an efficient class of generic algorithms (belonging to the class of derivative free methods) based on the surrogate management framework~\cite{Marsden2008} and artificial neural networks~\cite{Pierret2007} have been used for optimization in fluid mechanics, mainly in the area of aerodynamic shape optimization, they could require many function evaluations, for training purposes for example. When detailed simulations of interfacial flows are concerned, each function evaluation commands a full (potentially unsteady) CFD computation, causing gradient-based methods to be at an advantage. Adjoint-based algorithms present a suitable \changee{method}, as they allow the determination of the gradient at a cost comparable to a single function evaluation~\cite{Giles2000, griewank_algorithm_2000, rodrigues_adjoint-based_2020, burghardt_discrete_2022}. The use of adjoint methods for design and optimization has been an active area of research which started with the pioneering work of Pironneau~\cite{Pironneau1974} with applications in fluid mechanics, and has been extensively used in aeronautical shape optimization by Jameson and co-workers~\cite{Jameson88,Jameson98}. Ever since these groundbreaking studies, adjoint-based methods have been widely used in fluid mechanics particularly in the areas of aero- and thermo-acoustics~\cite{Juniper2010,Lemke2013}.
More recently flow regimes dominated by nonlinear dynamics, such as separation and mixing have also been analysed using adjoint-based techniques~\cite{Schmidt2013,Rabin2014,Foures2014,Duraisamy2012}. Adjoint-based methods have also been employed for the purpose of sensitivity analysis or control in flows in the presence of large gradients (flames or interfaces)~\cite{Fikl2020,ou_unsteady_2011,Braman2015,Lemke2019}, showing great promise, and therefore are adopted here to carry out the optimization  procedure.

In the context of Stefan problems various control strategies have been employed to track the location of the interface. In a one-dimensional setting, for example, set-valued fixed point equations~\cite{Hoffmann1982} or linear-quadratic defect minimization~\cite{Knabner1985} have been used to control the location of the front. Adjoint-based algorithms have also been applied to a Stefan problem caused by heterogeneous reactions on a surface of a one-dimensional solid particle~\cite{Hassan2021} to extract sensitives with respect to various kinetic parameters. Alternatively, in a two-dimensional setting, adjoint-based algorithms have been utilized previously together with finite element and finite difference approaches to track and control the location of interface by imposing heat flux (or temperature) at the boundary in order to realize the desired interface motion~\cite{Kang1995,Yang1997,Hinze2007}. In particular, Bernauer \& Herzog~\cite{bernauer_optimal_2011}, making use of shape calculus tools, derived the set of adjoint equations to extract control strategies for Stefan problems with a sharp representation of the interface. 

Similar to the approach of~\cite{bernauer_optimal_2011} shape calculus tools have been employed to extract the corresponding adjoint equations in our previous work~\cite{fullana2022}. However, contrary to the previous studies, control strategies are extracted here to control the interface shape resulting from the melting boundary. \changee{For the configuration considered here, deriving the fully coupled analytical adjoint of the Navier–Stokes–Stefan system on a moving domain is not tractable, which motivates the use of an incomplete continuous adjoint tailored to this problem. The} incomplete adjoint \changee{is derived} -- by considering the velocity field in the fluid layer as a \changee{known variable in the adjoint convection-diffusion equation} -- and used to minimize a tracking-type cost functional through a gradient-based algorithm. \changee{We will show that the derived incomplete adjoint is sufficient to retrieve meaningful information on the system and to drive the cost functional towards its minimum at a substantially reduced computational cost.}

\changee{The main contributions of this work can be summarized as follows. (i) A sharp-interface optimal-control problem for the Rayleigh–Bénard instability with a melting boundary, based on a level-set/cut-cell discretization of the two-phase Stefan problem under the Boussinesq approximation is formulated. (ii) We then derive an incomplete continuous adjoint that neglects the Navier–Stokes adjoint by treating the velocity field from the forward simulation as a known variable in the adjoint convection–diffusion equation. (iii) We demonstrate, on two non-trivial optimization problems and in comparison with a derivative-free method, that the resulting gradients are sufficiently accurate to control the front shape while significantly reducing the number of function evaluations. }

\changee{The present study is restricted to a two-dimensional configuration with a sharp interface representation and without surface-tension effects. The control is exerted only through the temperature distribution prescribed at the upper boundary of a periodic domain. As such, the proposed framework is intended as a first step towards adjoint-based control of melting fronts in convection-driven problems.}
  
The paper is organized as follows; in the first section we will define the forward problem and present validations cases, in the second section we will derive the continuous adjoint equations and in the last section we present the optimization results resulting from the solution of the minimization of a tracking-type cost functional.   

\section{Governing equations}\label{sec:equations}

\begin{figure}[t!]
\centering
\includegraphics[width=1\textwidth]{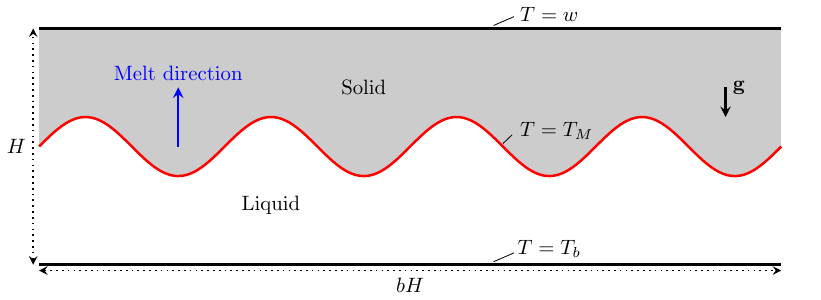}
\caption[Schematic of the melting boundary problem.]{Schematic of the melting boundary problem.}
\label{fig:schematic rayleigh}
\end{figure}

Following the recent studies on the Rayleigh-Bénard instability in the presence of a melting boundary \cite{Favier2019, Limare2022}, we consider the evolution of an initially flat liquid layer heated from below and comprised between a wall and a solid layer. As the solid melts, the liquid layer will grow vertically until a critical height where convection kicks in. A schematic of this configuration is shown in Fig.~\ref{fig:schematic rayleigh}. The gravity is pointing downwards $\mathbf{g} = -g \mathbf{e}_y$ and the horizontal size of the domain is $b H$, where $b$ is the aspect ratio and where $H = 1$ is the height of the system. The temperature at the bottom wall is $T_b = 0.7$, and the one at the top wall is our control variable $w(x) < 0$, such that $w(x) < T_M < T_b$ where $T_M = 0$ is the melting temperature at the front. \change{The choice of $T_b$ is such that $\Delta T = 1$ when $w(x) = -0.3$.} The initial flat interface is located at $h_0 = 0.05$ and the initial temperature field is set to zero.

\change{In the fluid phase, we solve the Navier-Stokes equations in the Boussinesq approximation coupled to the convection-diffusion equation for the temperature field. With this approximation, the variation of the liquid density with the temperature is taken into account
in the buoyancy force only. This means that the density and thermal conductivities are assumed to be constant and
equal. The equations read} 
\begin{equation} \label{eq:boussinesq}
\begin{split}
\operatorname{Pr}^{-1} \left( \dfrac{\partial \mathbf{u}}{\partial t} + \mathbf{u} \cdot \nabla \mathbf{u} \right) = & - \nabla p + \operatorname{Ra} \, T_L \, \mathbf{e}_y + \nabla^2 \mathbf{u}, \\
\nabla \cdot \mathbf{u} = & 0, \\
\dfrac{\partial T_L}{\partial t} + \mathbf{u} \cdot \nabla T_L = & \nabla^2 T_L,
\end{split}
\end{equation}
where $T_L$ is the dimensionless temperature field in the liquid phase. The dimensionless numbers governing this equation are the Prandtl number
\begin{equation} \label{eq:Prandlt number} 
\operatorname{Pr} = \dfrac{\nu}{k},
\end{equation}
defined by the ratio of the liquid kinematic viscosity $\nu$ and thermal diffusivity $k$, and the Rayleigh number
\begin{equation} \label{eq:Rayleight number} 
\operatorname{Ra} = \dfrac{g \, \alpha_t \Delta T \, H^3}{\nu \, k},
\end{equation}
where $g$ is the constant gravitational acceleration, $\alpha_t$ the coefficient of thermal expansion, $H$ a characteristic length of the system and $\Delta T = T_b - T_M$. \change{We apply no slip boundary conditions on the lower boundary and on the interface between the two phases.} In the solid phase, the heat equation applies
\begin{equation} \label{eq:boussinesq2} 
\begin{array}{rcl}
\dfrac{\partial T_S}{\partial t} &=& \nabla^2 T_S,
\end{array}
\end{equation}
where $T_S$ is the dimensionless temperature field in the solid phase. At the interface, both temperature fields are equal to the melting temperature
\begin{equation} \label{eq:melting}
T_L(\mathbf{x}, t) = T_S(\mathbf{x}, t)  = T_M \: \: \:\mathbf{x} \in \Gamma,
\end{equation}
where $\Gamma$ denotes the front position. In the present example, we disregard surface tension effects in the thermodynamic boundary condition at the front (ie. the Gibbs-Thomson relation). 

Finally, as phase change occurs, there will appear a latent heat which is either absorbed or released. The condition of heat conservation at a given point on the moving interface corresponds to the rate at which heat is generated at the boundary, balanced by the rate at which this heat flows in either phase. Along the interface, the Stefan condition~\cite{Stefan1891} states that
\begin{equation} \label{eq:stefan condition}
v  = \operatorname{St} \, [\nabla T]^S_L \cdot \mathbf{n}, \: \: \: \mathbf{x} \in \Gamma,
\end{equation}
where $\mathbf{v}$ is the interface velocity, $\mathbf{n}$ is the outward normal unit vector at the interface and $[\nabla T]^S_L = \partial T_{S} / \partial \mathbf{n} - \partial T_{L} / \partial \mathbf{n}$ is the jump in normal components of the temperature gradients from the solid phase to the liquid phase. The Stefan number $\operatorname{St}$ is defined as the ratio between available heat in the system and the latent heat  
\begin{equation} \label{eq:stefan_number}
\operatorname{St} = \dfrac{k \, \Delta T}{L_H},
\end{equation}
where $L_H$ is the latent heat. The moving interface is captured using the level set method. 
A level set function $\phi$ \cite{Sethian1999} is defined on the computational domain $\Omega$ to map the locus of one of its iso-levels ($\left \{ \left . \mathbf{x} \in \Omega \right | \phi \left ( \mathbf{x}, t \right ) = \phi _ 0 \right \}$) to an interface $\Gamma \left ( t \right )$ that separates two non-overlapping domains, $\Omega _ S \left ( t \right )$ and $\Omega _ L \left ( t \right )$, each occupied by a different phase. The value $\phi$ is defined as the signed distance to the interface,
\begin{equation}
	\phi \left ( \mathbf{x}, t \right ) = \left \{ \begin{aligned}
		-d \left ( \mathbf{x}, \Gamma \left ( t \right ) \right ),& \ \mathbf{x} \in \Omega _ S \left ( t \right ) \\
		0,& \ \mathbf{x} \in \Gamma \left ( t \right ) \\
		d \left ( \mathbf{x}, \Gamma \left ( t \right ) \right ),& \ \mathbf{x} \in \Omega _ L  \left ( t \right )
	\end{aligned} \right .,
\end{equation}
where $d \left ( \mathbf{x}, \Gamma \left ( t \right ) \right )$ denotes the minimal distance between the point $\mathbf{x}$ and the interface $\Gamma \left ( t \right )$,
\begin{equation}
	d \left ( \mathbf{x}, \Gamma \left ( t \right ) \right ) = \operatorname{argmin} _ {\mathbf{y} \in \Gamma \left ( t \right )} \left \Vert \mathbf{x} - \mathbf{y} \right \Vert ,
\end{equation}
with $\left \Vert \cdot \right \Vert$ denoting the Euclidean distance. 
The main advantage of the level set method is that its evolution is governed by a simple advection equation
\begin{equation} \label{eq:level set_advection}
    \pder{\phi}{t} + v \cdot \nabla \phi = 0.
\end{equation}
provided that the velocity field can be smoothly extended from $\Gamma \left ( t \right )$. \change{The most natural algorithm is to let $v$ be a constant along the lines normal to $\Gamma$. To achieve this, the method described in~\cite{peng_pde-based_1999} is adopted here. Using this approach, the velocity is extended in the normal direction by solving the following hyperbolic partial differential equation
\begin{equation}\label{eq:velo_extension}
\left\{\begin{array}{rr}
\dfrac{\partial F}{\partial t^\star}+S(\phi) \dfrac{\nabla \phi}{|\nabla \phi|} \cdot \nabla F=0&\text{in} \: \Omega \\
F(x, 0)=v & \text{on} \: \Gamma
\end{array}
\right.
\end{equation} 
where $F$ is the extended velocity field equal to $v$ at the front, $t^\star$ denotes a pseudo-time and $S(\phi)$ is the signature function 
\begin{equation}\label{eq:signature}	
S(\phi)=\left\{\begin{array}{ll}
-1 & \text { if } \phi<0 \\
0 & \text { if } \phi=0 \\
+1 & \text { if } \phi>0
\end{array}\right.
\end{equation} 
Equation~\ref{eq:velo_extension} is then discretized using a first order upwind scheme and integrated in time by a forward Euler method until steady state. Taking $\mathbf{n}$ as the normal vector defined as
\begin{equation}\label{eq:normal_vector}
n = (n_x, n_y)  = \left(\phi_x / \sqrt{(\phi_x^2 + \phi_y^2)}, \phi_y / \sqrt{(\phi_x^2 + \phi_y^2)} \right),
\end{equation}
the discretisation leads to
\begin{equation}\label{eq:discrete_velo}
\begin{aligned}
F_{i j}^{n+1}=& F_{i j}^{n}-\tau^\star \left(\left(S_{i j} n_{i j}^{x}\right)^{+} \displaystyle \dfrac{F_{i j}-F_{i-1 j}}{\Delta}+\left(S_{i j} n_{i j}^{x}\right)^{-} \dfrac{F_{i+1 j}-F_{i j}}{\Delta}\right.\\
&\left.+\left(S_{i j} n_{i j}^{y}\right)^+\displaystyle \dfrac{F_{i j}-F_{i j-1}}{\Delta}+\left(S_{i j} n_{i j}^{y}\right)^-\dfrac{F_{i j+1}-F_{i j}}{\Delta}\right)
\end{aligned}
\end{equation}
where $\Delta$ is the uniform grid spacing, $(x)^+ = \operatorname{max}(0,x)$ and $(x)^- = \operatorname{min}(0,x)$, and the time step $\tau^\star$ is chosen so that $\tau^\star / \Delta^2 = 0.45$. The pseudo-time spawn in the velocity extension algorithm is purely fictitious and the number of iterations in Equation~\ref{eq:discrete_velo} corresponds to the width of the narrow-band (NB) around the 0-level set where the velocities are initialized. 
A semi-implicit scheme described in~\cite{mikula_inflow-implicitoutflow-explicit_2014, mikula_new_2010} is used here to solve the level set advection equation. Using this method, the equation is written in an alternative form
\begin{equation}\label{eq:mikula0}
    \dfrac{\partial \phi}{\partial t} + F \dfrac{\nabla \phi}{|\nabla \phi|} \cdot \nabla \phi = 0,
\end{equation}
which is then divided into conservative and non-conservative terms
\begin{equation}\label{eq:mikula1}
    \dfrac{\partial \phi}{\partial t} + \underbrace{\nabla \cdot \left( F \phi \dfrac{\nabla \phi}{|\nabla \phi|} \right)}_{a} - \underbrace{\phi \nabla \cdot \left( F \dfrac{\nabla \phi}{|\nabla \phi|} \right)}_{b} = 0,
\end{equation} 
resulting in a second order partial differential equation akin to a weighted diffusion equation. The first term ($a$) has a diffusion coefficient $F \phi$ that depends on the solution and represents a nonlinear curvature flow whereas in the second term ($b$) the solution is multiplied by the curvature of its level-sets. The main idea behind this scheme is to distinguish two cases: if the product $F \phi$ is negative (positive, respectively) then $a$ represents a forward (backward, respectively) diffusion and $b$ represents a backward (forward, respectively) diffusion. The forward diffusion is treated implicitly while the backward diffusion is treated explicitly leading to a semi-implicit scheme with a diffusive CFL number (fixed to 0.5 in the test cases presented here).\\
In order to discretize Equation~\ref{eq:mikula1}, we use the same notation as in~\cite{mikula_inflow-implicitoutflow-explicit_2014}. Consider $\mathrm{p}$ to be a finite volume of a cell and $e_{\mathrm{p}\mathrm{q}}$ the edge between $\mathrm{p}$ and $\mathrm{q}$, $\mathrm{q} \in \mathcal{N}(\mathrm{p})$ where $\mathcal{N}(\mathrm{p})$ is the set of neighboring finite volumes. The length of $e_{\mathrm{p}\mathrm{q}}$ is normalized to $1$. Let $\mathbf{n}_{\mathrm{p}\mathrm{q}}$ be the outer normal vector to $e_{\mathrm{p}\mathrm{q}}$ with respect to $\mathrm{p}$. Finally, let us denote $\bar{\phi}_\mathrm{p}$ the constant reconstruction of $\phi_\mathrm{p}$ in the finite volume $p$ and $\bar{\phi}_{\mathrm{p}\mathrm{q}}$ the constant reconstruction of $\phi$ on $e_{\mathrm{p}\mathrm{q}}$. Integrating volume $\mathrm{p}$, Equation~\ref{eq:mikula1} yields
\begin{equation}\label{eq:mikula2}
\int_{\mathrm{p}} \dfrac{\partial \phi}{\partial t} dx + \int_{\mathrm{p}} \nabla \cdot \left( F \phi \dfrac{\nabla \phi}{|\nabla \phi|} \right) dx - \int_{\mathrm{p}} \phi \nabla \cdot \left( F \dfrac{\nabla \phi}{|\nabla \phi|} \right) dx = 0.
\end{equation}
Applying Stokes' theorem and using the constant reconstructions of $\phi$, we obtain
\begin{align*}
\int_{\mathrm{p}} \dfrac{\partial \phi}{\partial t} dx + \sum_{\mathrm{q} \in \mathcal{N}(\mathrm{p})} \bar{\phi}_{\mathrm{p}\mathrm{q}}\int_{e_{\mathrm{p}\mathrm{q}}} F \dfrac{1}{|\nabla \phi|} \nabla \phi \cdot \mathbf{n}_{\mathrm{p}\mathrm{q}} ds\\
-  \sum_{\mathrm{q} \in \mathcal{N}(\mathrm{p})} \bar{\phi}_{\mathrm{p}} \int_{e_{\mathrm{p}\mathrm{q}}} F \phi \dfrac{1}{|\nabla \phi|} \nabla \phi \cdot \mathbf{n}_{\mathrm{p}\mathrm{q}} ds = 0.
\end{align*}
Let us denote $|\nabla \phi_{\mathrm{p}\mathrm{q}}|$ the reconstructed Hamiltonian $|\nabla \phi|$ on the edge $e_{\mathrm{p}\mathrm{q}}$ and $\left( \phi_{\mathrm{q}} - \phi_{\mathrm{p}} \right)/1$ the normal derivative $\nabla \phi \cdot \mathbf{n}_{\mathrm{p}\mathrm{q}}$ on the same edge. We obtain the following expression
\begin{align*}\label{eq:mikula4}
\int_{\mathrm{p}} \dfrac{\partial \phi}{\partial t} dx + \sum_{\mathrm{q} \in \mathcal{N}(\mathrm{p})} \dfrac{F \bar{\phi}_{\mathrm{p}\mathrm{q}}}{|\nabla \phi_{\mathrm{p}\mathrm{q}}|} (\phi_\mathrm{q} - \phi_\mathrm{p})\\
- \sum_{\mathrm{q} \in \mathcal{N}(\mathrm{p})} \dfrac{F \bar{\phi}_{\mathrm{p}}}{|\nabla \phi_{\mathrm{p}\mathrm{q}}|} (\phi_\mathrm{q} - \phi_\mathrm{p})= 0,
\end{align*}
leading to
\begin{equation}\label{eq:mikula5}
\int_{\mathrm{p}} \dfrac{\partial \phi}{\partial t} dx + \sum_{\mathrm{q} \in \mathcal{N}(\mathrm{p})} \dfrac{F (\bar{\phi}_\mathrm{p} - \bar{\phi}_{\mathrm{p}\mathrm{q}})}{|\nabla \phi_{\mathrm{p}\mathrm{q}}|} (\phi_\mathrm{p} - \phi_\mathrm{q}) = 0.
\end{equation}
Looking at the term $F (\bar{\phi}_\mathrm{p} - \bar{\phi}_{\mathrm{p}\mathrm{q}})$, we can distinguish two cases: 
\begin{itemize}
\item If the term is positive, we have a ‘forward diffusion’ or inflow towards the cell.
\item If the term is negative, we have a ‘backward diffusion’ or outflow from the cell.
\end{itemize}
We therefore define the diffusion coefficient $a_{\mathrm{p}\mathrm{q}}$ as 
\begin{equation}\label{eq:mikula6}
a_{\mathrm{p}\mathrm{q}} = \dfrac{F (\bar{\phi}_\mathrm{p} - \bar{\phi}_{\mathrm{p}\mathrm{q}})}{|\nabla \phi_{\mathrm{p}\mathrm{q}}|},
\end{equation}
and the related dominant forward and backward diffusion parts as
\begin{equation}\label{eq:mikula6b}
a^f_{\mathrm{p}\mathrm{q}} =  \operatorname{max}(a_{\mathrm{p}\mathrm{q}}, 0), \quad \quad a^b_{\mathrm{p}\mathrm{q}} =  \operatorname{min}(a_{\mathrm{p}\mathrm{q}}, 0).
\end{equation}
Using a backward Euler time discretization, taking the forward contribution explicitly and the backward contribution implicitly, Equation~\ref{eq:mikula5} gives the following linear system 
\begin{equation}\label{eq:mikula7}
\underbrace{\phi^n_\mathrm{p} + \dfrac{\tau}{\Delta^2} \sum_{\mathrm{q} \in \mathcal{N}(\mathrm{p})} a^f_{\mathrm{p}\mathrm{q}} (\phi^n_\mathrm{p} - \phi^n_\mathrm{q})}_{\operatorname{implicit}} = \underbrace{\phi^{n - 1}_\mathrm{p} + \dfrac{\tau}{\Delta^2} \sum_{\mathrm{q} \in \mathcal{N}(\mathrm{p})} a^b_{\mathrm{p}\mathrm{q}} (\phi^{n - 1}_\mathrm{p} - \phi^{n - 1}_\mathrm{q})}_{\operatorname{explicit}}
\end{equation}
where $\tau$ is the time step, $\Delta$ the uniform grid spacing and $n$ a given time step. To reconstruct of $\bar{\phi}_\mathrm{p}$, $\bar{\phi}_{\mathrm{p}\mathrm{q}}$ and $|\nabla \phi_{\mathrm{p}\mathrm{q}}|$, we use the diamond-cell strategy described in~\cite{mikula_new_2010}. This evolution equation however does not preserve the signed distance property, which can be restored by periodically employing a reinitialization procedure by solving an Eikonal equation. There exist many numerical methods for solving this equation and, here we have adopted that of Min~\cite{min_reinitializing_2010}, also recently used in the work of Limare~\cite{Limare2022}.}
For further details on the numerical implementation of the level set coupled to a cut cell method, we refer the reader to~\cite{fullana2022, Fullana2023, QuirosRodriguez2024}. 

The forward problem, setting $\operatorname{St} = 1$, can be recast in the level set framework as follows.
\begin{center}
Find a function $T : \Omega \times[0, t_{f}] \rightarrow \mathbb{R}$ and a function $\phi : \Omega \times[0, t_{f}] \rightarrow \mathbb{R}$ such that
\[\arraycolsep=2pt\def\arraystretch{2.2}
\label{FP} \tag{FP}\left\{
\begin{array}{rcllr} \dfrac{\partial T_S}{\partial t} &=& \nabla^2 T_S & \: \text{in} \: \Omega_S(t) & \text{(FP.a)}\\
\operatorname{Pr}^{-1} \left( \dfrac{\partial \mathbf{u}}{\partial t} + \mathbf{u} \cdot \nabla \mathbf{u} \right)& = & - \nabla p + \operatorname{Ra} \, T_L \, \mathbf{e}_y + \nabla^2 \mathbf{u} & \: \text{in} \: \Omega_L(t) & \text{(FP.b)}\\
\nabla \cdot \mathbf{u} & = & 0 & \: \text{in} \: \Omega_L(t) & \text{(FP.c)}\\
\dfrac{\partial T_L}{\partial t} + \mathbf{u} \cdot \nabla T_L & = & \nabla^2 T_L & \: \text{in} \: \Omega_L(t) & \text{(FP.d)}\\
\dfrac{\partial T_L}{\partial t} &=& \nabla^2 T_L & \: \text{in} \: \Omega_L(t) & \text{(FP.e)}\\ 
T(x,0) &=& T_0 \left ( x \right ) & \: \text{in}  \: \Omega & \text{(FP.f)}\\
\displaystyle T_S &=& w \left ( x \right) & \: \text{on} \: \partial \Omega_S & \text{(FP.g)}\\
\displaystyle T_L &=& T_b & \: \text{on} \: \partial \Omega_L & \text{(FP.h)}\\
T(x,t) &=& T_M & \: \text{on} \: \Gamma(t) & \text{(FP.i)}\\
\dfrac{\partial \phi}{\partial t} &=& -[\nabla T_i]^S_L \cdot \nabla \phi & \: \text{on} \: \Gamma(t) & \text{(FP.j)}\\
\phi(x,0) &=& \phi_0 \left ( x \right ) & \: \text{in} \: \Omega & \text{(FP.k)}
\end{array}
\right.
\]
\end{center}

\begin{figure}[ht!]
\centering
\subfloat[Average height]{
\centering
\includegraphics[width=0.47\linewidth]{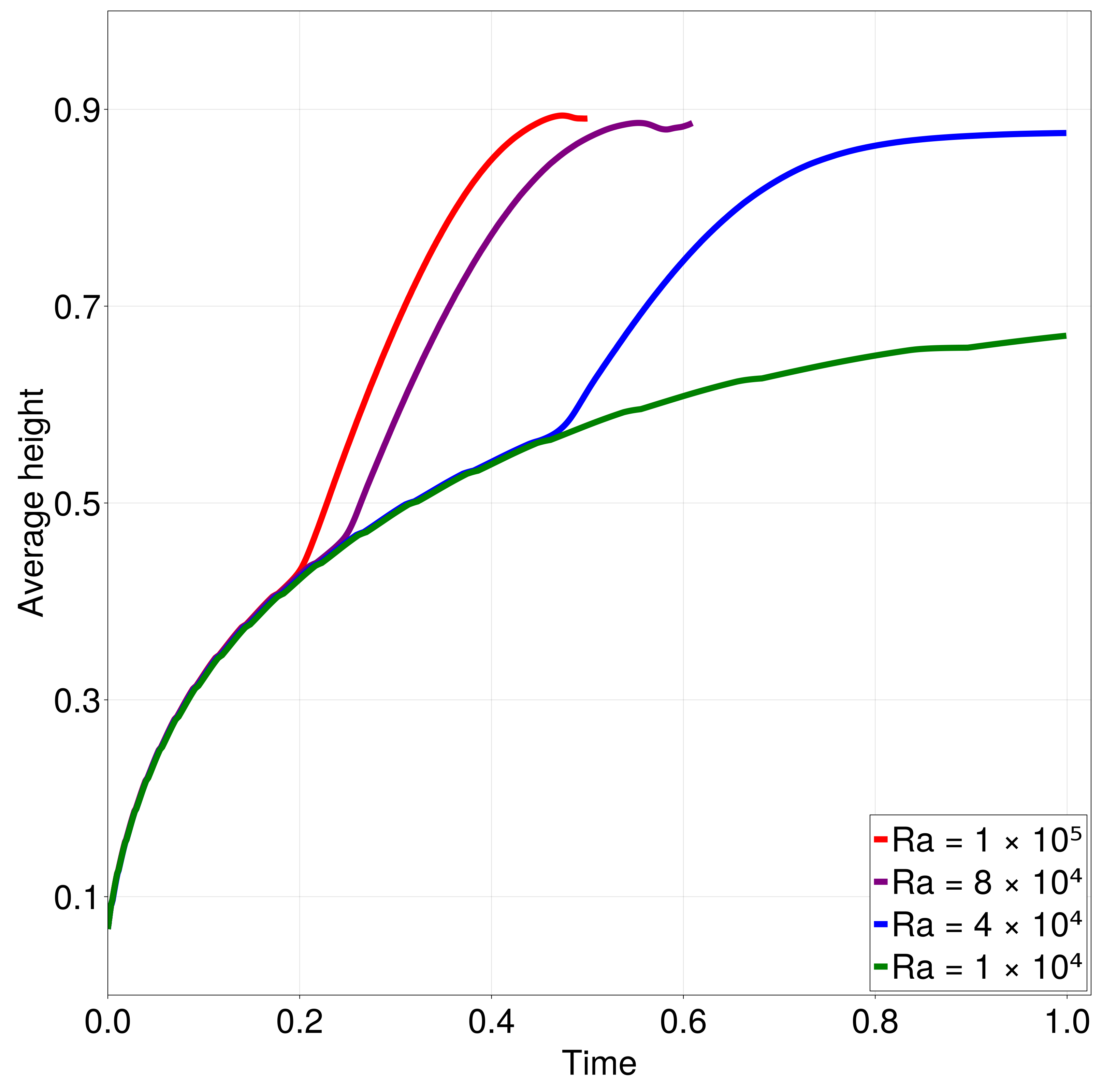}
}
\subfloat[Effective Rayleigh number]{
\centering
\includegraphics[width=0.47\linewidth]{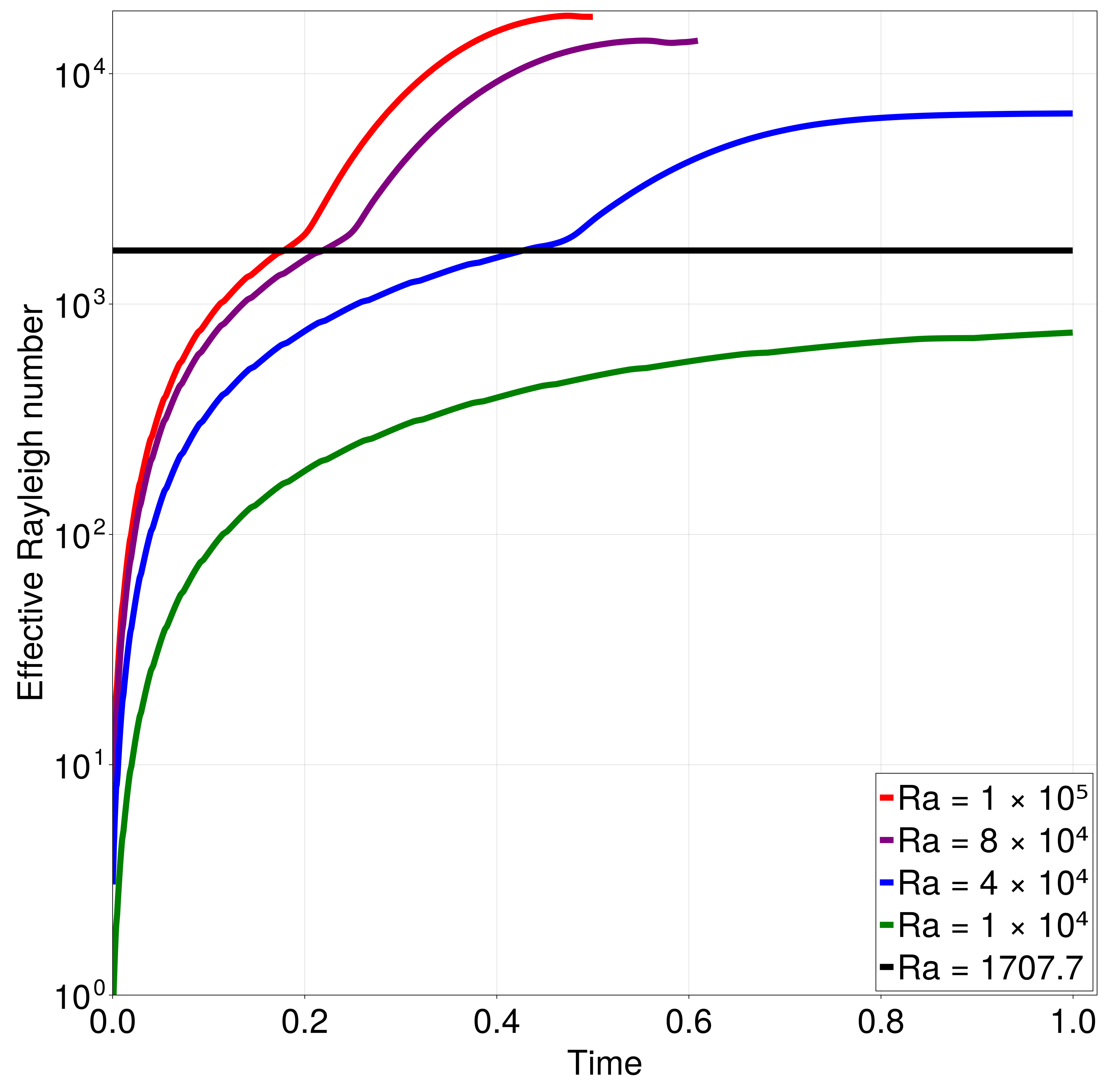}
}
    \caption{Average height and effective Rayleigh number as a function of time for $\operatorname{Ra} = 10^5, \, 8 \times10^4, \, 4 \times 10^4, \, 10^4.$}
    \label{fig:hRa}
\end{figure}

\begin{figure}[ht!]
\centering
\subfloat[t = 0.1]{
\centering
\includegraphics[width=1\linewidth]{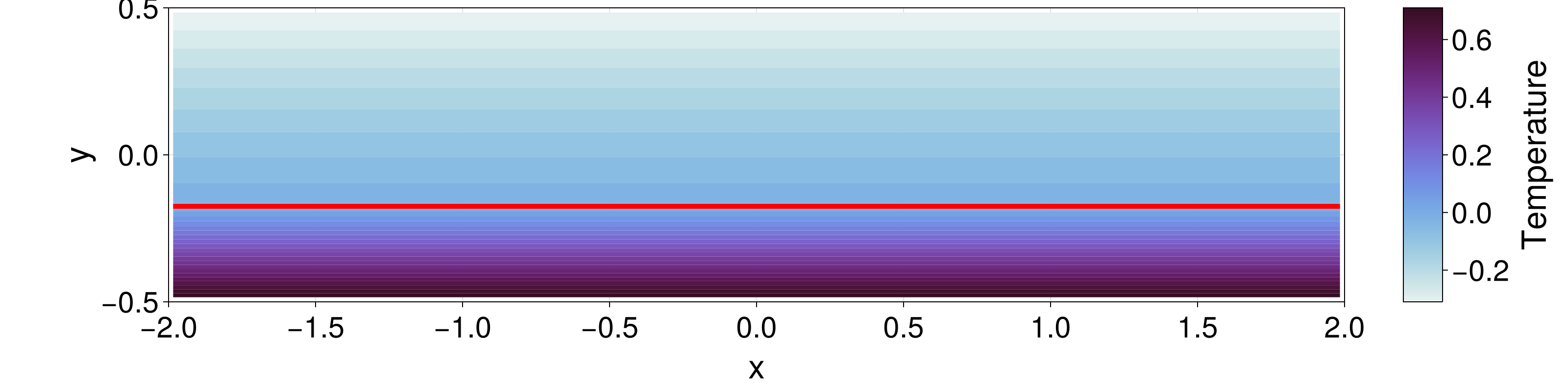}
}

\subfloat[t = 0.2]{
\centering
\includegraphics[width=1\linewidth]{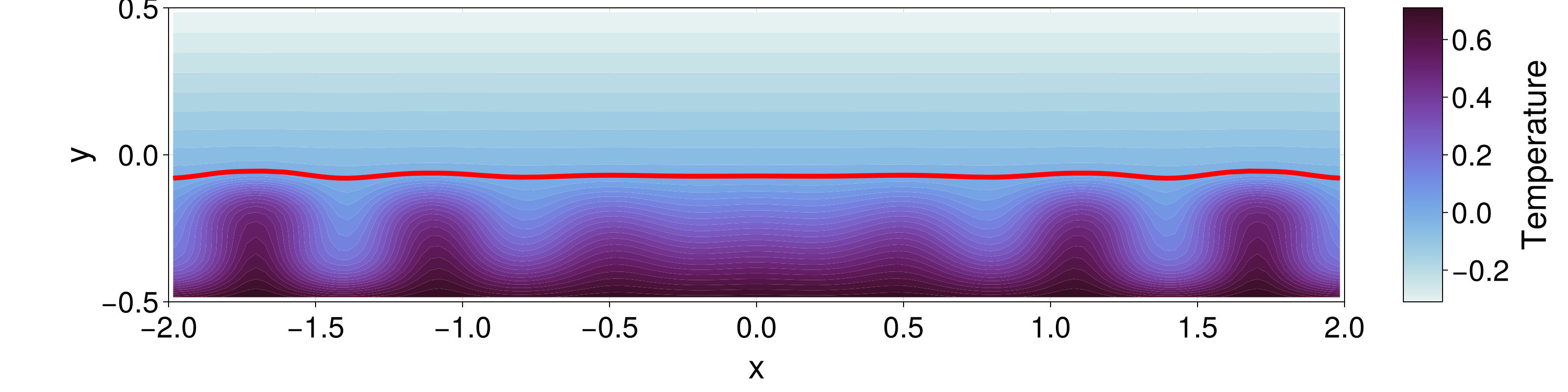}
}

\subfloat[t = 0.3]{
\centering
\includegraphics[width=1\linewidth]{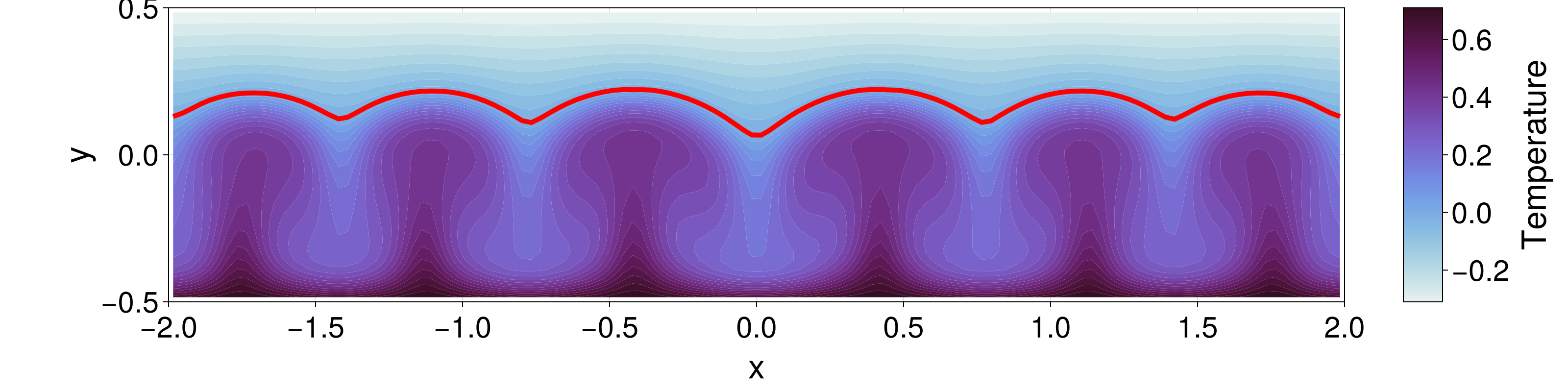}
}

\subfloat[t = 0.4]{
\centering
\includegraphics[width=1\linewidth]{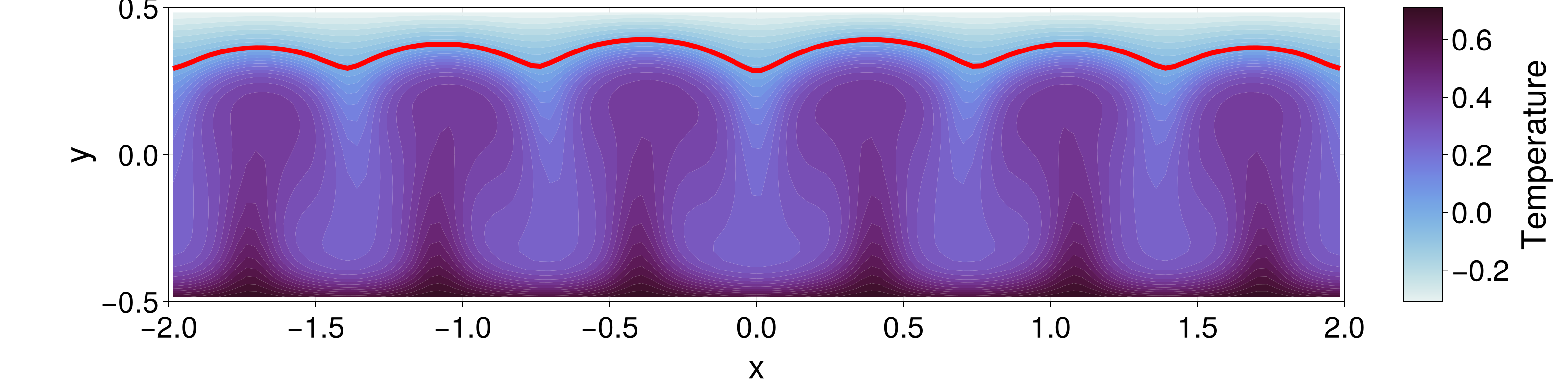}
}
    \caption{Times series of the temperature field and interface position for the $\operatorname{Ra} = 10^5$ case. The color map is the temperature field in both phases and the interface corresponding to the 0-level set is denoted in red.}
    \label{fig:RB_fwd}
\end{figure}

\begin{figure}[ht!]
\centering
\subfloat[t = 0.1]{
\centering
\includegraphics[width=1\linewidth]{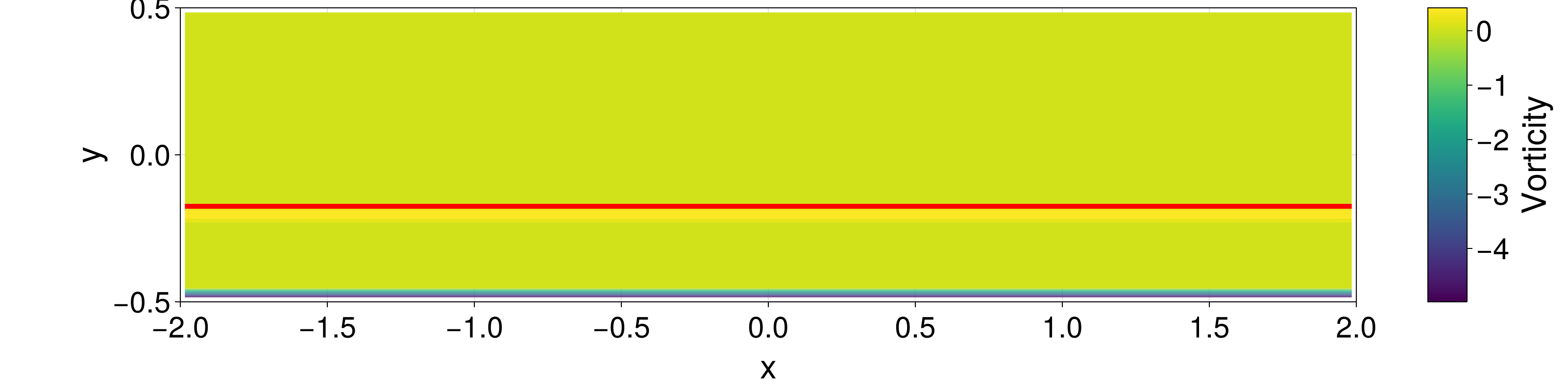}
}

\subfloat[t = 0.2]{
\centering
\includegraphics[width=1\linewidth]{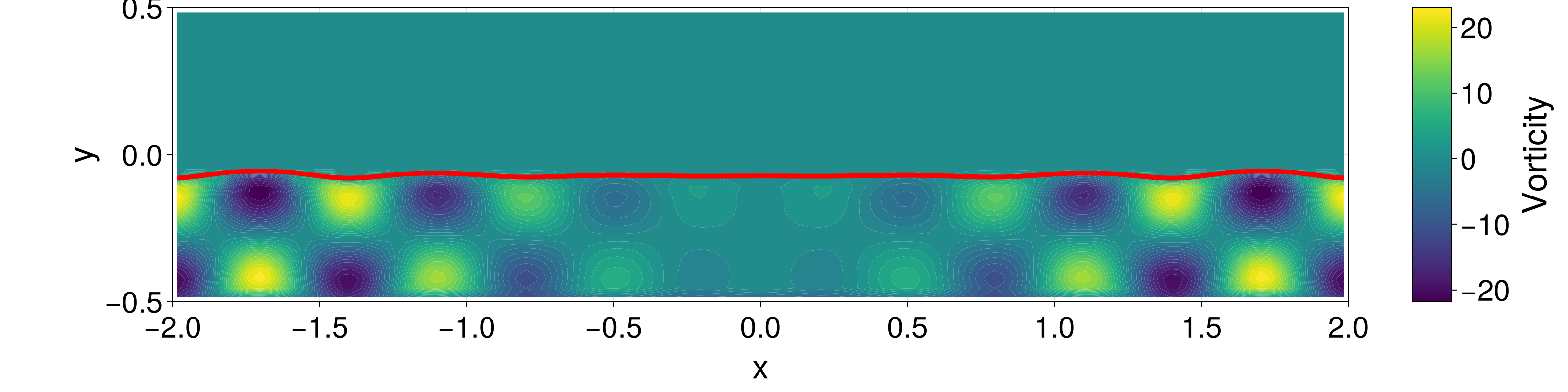}
}

\subfloat[t = 0.3]{
\centering
\includegraphics[width=1\linewidth]{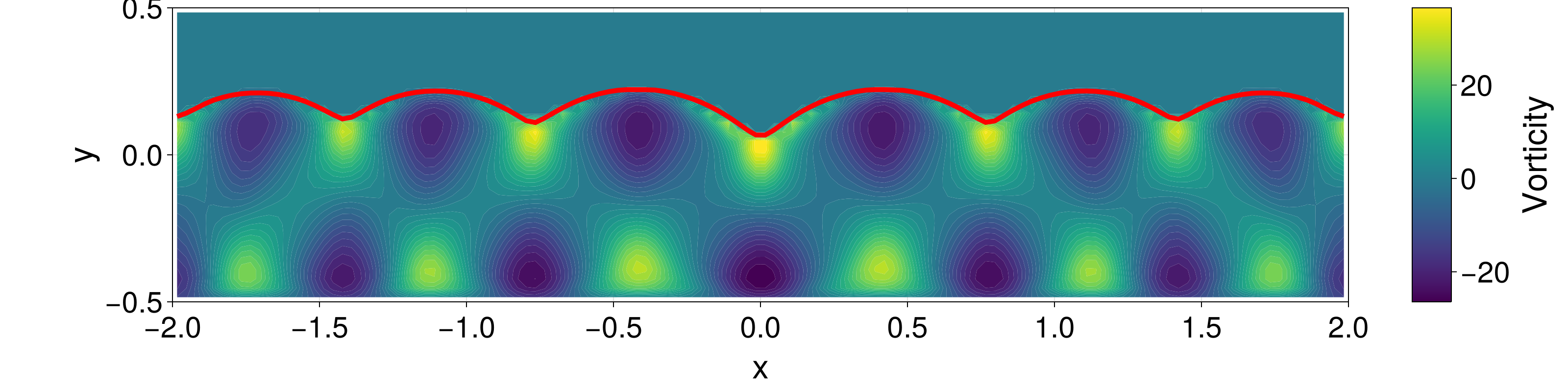}
}

\subfloat[t = 0.4]{
\centering
\includegraphics[width=1\linewidth]{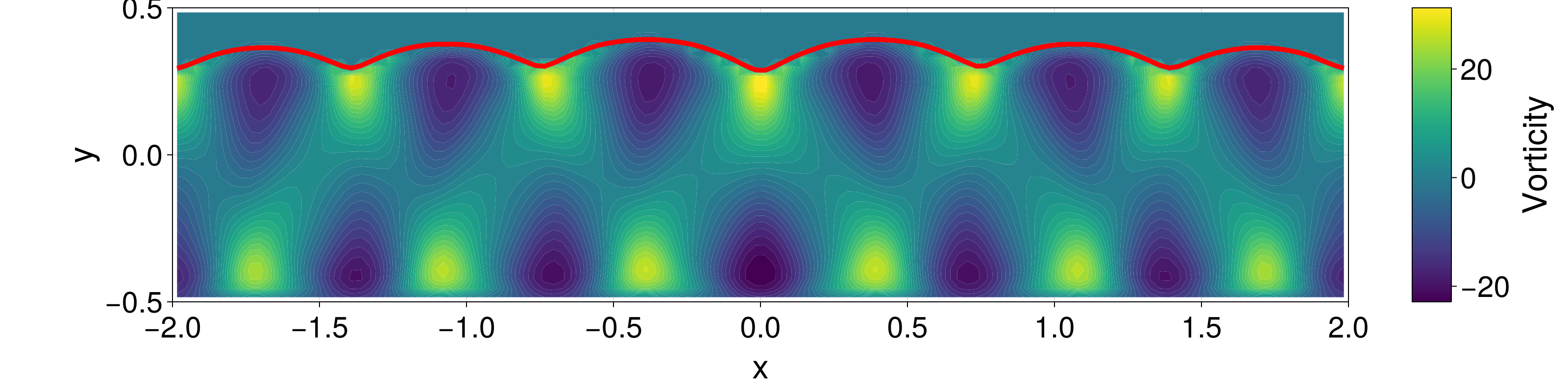}
}
    \caption{\change{Times series of the vorticity and interface position for the $\operatorname{Ra} = 10^5$ case. The interface corresponding to the 0-level set is denoted in red.}}
    \label{fig:vort_fwd}
\end{figure}

The global Rayleigh number in Eq.~\eqref{eq:Rayleight number} will control the onset of the Rayleigh-Bénard instability. The effective Rayleigh number is defined as
\begin{equation}\label{eq:Rae}
\operatorname{Ra}_e = \operatorname{Ra} \, (1 - T_M) \, \bar{h}^3,
\end{equation}
where $\bar{h}$ is the average fluid height 
\begin{equation}\label{eq:avh}
\bar{h}(t) = \dfrac{1}{b} \int_0^b h(x, t) dx.
\end{equation}
When the effective Rayleigh number reaches the critical value $\operatorname{Ra}_c = 1707.76$, the initial diffusion-driven motion is transformed into a convection-driven one.
We consider a domain with aspect ratio $b = 4$, a fixed temperature $w = -0.3$ at the top boundary and vary the global Rayleigh number. Figure~\ref{fig:hRa}a shows the average height as function of time for the different cases. In the $\operatorname{Ra} = 10^4$ case, the critical Rayleigh number is not reached and the motion of the fluid layer is not affected, thus remaining a diffusion-driven one, similarly to the planar motion. In the rest of the cases, the effective Rayleigh number $\operatorname{Ra}_e$ (Fig.~\ref{fig:hRa}b) crosses the threshold indicating the onset of the instability, characterized by a sharp increase in interface speed. Figure~\ref{fig:RB_fwd} shows a time series of the temperature field and interface position for the $\operatorname{Ra} = 10^5$ case  \change{and Figure~\ref{fig:vort_fwd} shows the vorticity field for the same case}. When the critical Rayleigh number is reached, the first bifurcation appears, creating the convection cells. The size of the convection cells will then vary with the secondary bifurcations mechanism. When the averaged height $\bar{h}$ matches the characteristic wavelength of the convection rolls, the convection cells have sufficient time to merge and then stabilize. We also note that the interface is deformed according to the shape of the cells.

\FloatBarrier
\section{Adjoint problem}\label{sec:methods}

We now aim at solving a minimization problem by using the adjoint of \eqref{FP}. We assume that the desired temperature field $T^d$ and the desired position of the interface $\phi^d$ are known a priori. The control variable $w$ is then iteratively calculated trough an optimization procedure to reach these desired quantities and will drive our cost functional towards its minimum. The following tracking-type cost functional provides a mathematical description of the control goals stated above
\begin{equation}
    \label{eq:cost_functional}
    \mathcal{J} (T, \phi, w) =
\displaystyle \frac{\beta_{1}}{2} \int_{\Omega}\left|T^f-T^d\right|^{2} dx +\frac{\beta_{2}}{2} \int_{\Gamma^f}\left|\phi^f-\phi^d\right|^{2} ds +\frac{\beta_{3}}{2} \int_{0}^{t_{f}} \int_{\partial \Omega_S}|w|^{2} ds \: dt
\end{equation}
with constants $\beta_1$ to $\beta_3$ acting as weights, $T^f$ and $\phi^f$ the final temperature field and level set function of~\eqref{FP} and $\Gamma^f$ the final position of the interface (0-level set of $\phi^f$). The first term monitors the final temperature distribution and is mostly used as an initializer for the adjoint temperature field for the adjoint problem. The second term controls the relative position of the level set function with respect to the desired one. Taking advantage of the signed distance function property of both $\phi^d$ and $\phi^f$, the discrete form is simply the square of the difference between both functions computed in the points belonging to the final interface $\Gamma^f$. Finally, the last term penalizes the control cost and leads to the gradient equation in the adjoint problem.

We assume that each choice of the control variable $w$ leads to unique states $T(w)$ and $\phi(w)$. Therefore, the minimization problem \eqref{MP} reads
\begin{equation}
\label{MP} \tag{MP}
\begin{array}{c}
\operatorname{min} _{w} \mathcal{J} (T(w), \phi(w), w) \\
\text{subject to \eqref{FP}}.
\end{array}
\end{equation}


\changee{We will now proceed with the derivation of the corresponding continuous adjoint problem. The main novelty behind our derivation is that the velocity field $\mathbf{u}$, known at each time step from the forward solution, is treated as a known variable in the adjoint equations. This allows us to discard the adjoint of the Navier–Stokes equations in the fluid phase and leads to an incomplete adjoint formulation.} The minimization procedure will serve as a validation of this hypothesis.

Let $\Theta$ be the adjoint temperature and $\psi$ the adjoint level set function. In order to compute the gradient of $\mathcal{J}(w)$, we introduce the following Lagrange functional $\mathcal{L}$ ($dx$, $dt$ and $ds$ are omitted for brevity)
\begin{equation}\label{eq:Lagrange}
\arraycolsep=2pt\def\arraystretch{2.2}
\begin{array}{l}
\mathcal{L}\left(T, \phi, w, \Theta, \Theta_S^D, \Theta_L^D, \Theta^I, \psi\right)=\mathcal{J} (T, \phi, w)  \\
- \displaystyle \int_{0}^{t_{f}} \int_{\Omega_S} \left( \dfrac{\partial T_{S}}{\partial t} - \nabla^2 T_{S} \right) \Theta_S 
- \int_{0}^{t_{f}} \int_{\Omega_L} \left( \dfrac{\partial T_L}{\partial t} + \mathbf{u} \cdot \nabla T_L -\nabla^2 T_L\right) \Theta_L \\
- \displaystyle \int_{0}^{t_{f}} \int_{\partial \Omega_S} \left( T - w \right) \Theta_S^D 
- \displaystyle \int_{0}^{t_{f}} \int_{\partial \Omega_L} \left( T - T_b \right) \Theta_L^D 
- \displaystyle \int_{0}^{t_{f}} \int_\Gamma \left( T - T_M \right) \Theta^I \\
- \displaystyle \int_{0}^{t_{f}} \int_\Gamma \left(\dfrac{\partial \phi}{\partial t} + [\nabla T_i]^S_L \cdot \nabla \phi  \right) \psi.
\end{array}
\end{equation}
Similarly to the temperature field, $\Theta_{S, L} \colon \left (x, t \right ) \to \mathbb{R} ^ +$ denotes the adjoint temperature field in either phase. The Lagrange multipliers $\Theta^D_{S, L}$ and $\Theta^I$ are used for the boundary conditions on the domain and the interface respectively.

The adjoint system is obtained by setting to zero the derivatives of the Lagrange functional with respect to $T$ and $\phi$ : $\mathcal{L}_T(\cdot) = \mathcal{L}_\phi(\cdot) = 0$. Setting the initial conditions to
\begin{equation}
\begin{aligned}
T(x, 0) = T_0(x),\\
\phi(x, 0) = \phi_0(x),
\end{aligned}
\end{equation}
leads to $\delta T(x, 0) = \delta \phi (x, 0) = 0$ when calculating the derivatives in the direction $\delta T$ and $\delta \phi$.

We start by deriving the adjoint temperature equations by setting $\mathcal{L}_T(\cdot) = 0$. The terms in the Lagrangian that do not depend on the temperature vanish and we move the spatial and temporal derivatives towards the adjoint state $\Theta$. By applying integration by parts, once with respect to time and twice with respect to space, we obtain

\begin{equation*}\label{eq:LT}
\arraycolsep=2pt\def\arraystretch{2.2}
\begin{array}{l}
\mathcal{L}_T \delta T = \displaystyle \int_{0}^{t_{f}} \int_{\Omega_S} \left( \dfrac{\partial \Theta_S}{\partial t} + \nabla^2 \Theta_S \right) \delta T 
+ \displaystyle \int_{0}^{t_{f}} \int_{\Omega_L} \left( \dfrac{\partial \Theta_L}{\partial t} - \mathbf{u} \nabla \Theta_L + \nabla^2 \Theta_L \right) \delta T \\
- \displaystyle \int_{\Omega_S} \left(\Theta_S ^f - \beta_1 \left( T^f - T^d\right) \right) \delta T^f 
- \displaystyle \int_{\Omega_L} \left(\Theta_L ^f - \beta_1 \left( T^f - T^d\right) \right) \delta T^f \\
+ \displaystyle \int_{0}^{t_{f}} \int_{\partial \Omega_S} \left(- \dfrac{\partial \Theta_S}{\partial n} \delta T + \dfrac{\partial \delta T}{\partial n} \left(\Theta_S - \Theta_S^D\right) \right) + \displaystyle \int_{0}^{t_{f}} \int_{\partial \Omega_L} \left(- \dfrac{\partial \Theta_L}{\partial n} \delta T + \dfrac{\partial \delta T}{\partial n} \left(\Theta_L - \Theta_L^D\right) \right)\\
+ \displaystyle \int_{0}^{t_{f}} \int_{\Gamma} \left(\Theta_S \vec{v}_S \cdot n - \Theta_L \vec{v}_L \cdot n - \dfrac{\partial \Theta_S}{\partial n} + \dfrac{\partial \Theta_S}{\partial n} - \Theta^I \right) \delta T \\
+ \displaystyle \int_{0}^{t_{f}} \int_{\Gamma} \left( \Theta - \psi |\nabla \phi|\right) [\nabla \delta T]^S_L \cdot n \quad \quad \quad \quad \forall \delta T = 0.
\end{array}
\end{equation*}
with $\vec{v}_S$ and $\vec{v}_L$ the velocities of the control volumes $\Omega_S$ and $\Omega_L$ respectively, only non-zero on $\Gamma$. The second to last term, corresponds to the normal jump in adjoint temperature, which gives
$\Theta^I = -[\nabla \Theta]^S_L \cdot n$ as $\Theta_S = \Theta_L$ on $\Gamma$. By altering the directions of variations and eliminating certain terms, we obtain the adjoint temperature problem
\begin{center} 
\[\arraycolsep=2pt\def\arraystretch{2.2}
\label{AT} \tag{AT} \left\{
\begin{array}{rcllr} - \dfrac{\partial \Theta_{S}}{\partial t} &=& \nabla^2 \Theta_{S} & \: \text{in} \: \Omega_S(t) \\
- \dfrac{\partial \Theta_{L}}{\partial t} - \mathbf{u} \, \nabla \Theta_L &=& \nabla^2 \Theta_{L} & \: \text{in} \: \Omega_L(t) \\ 
\Theta(x,t_f) &=& \beta_1 (T^f-T^d) & \: \text{in}  \: \Omega \\
\dfrac{\partial \Theta_S}{\partial n} &=& 0 & \: \text{on} \: \partial \Omega_S \\
\dfrac{\partial \Theta_L}{\partial n} &=& 0 & \: \text{on} \: \partial \Omega_L \\
\Theta(x,t) &=& \psi |\nabla \phi| & \: \text{on} \: \Gamma(t) \\
\Theta^I &=& -[\nabla \Theta]^S_L \cdot n & \: \text{on} \: \Gamma(t)
\end{array}
\right.
\]
\end{center}
The first two equations of the adjoint temperature problem \eqref{AT} are the heat and convection-diffusion equations in reverse time. The third equation is the initial condition for the adjoint temperature field, that depends on the desired temperature distribution $T^d$. The fourth and fifth equations are the homogeneous Neumann boundary conditions for the adjoint field mapped from the Dirichlet boundary conditions of the forward problem. The second to last equation is the Dirichlet boundary condition at the interface that now depends on the value of the adjoint level set function $\psi$. We can note that the adjoint level set function will not behave as a signed distance function but as an auxiliary variable that acts on the temperature field through the boundary condition at the interface $\Gamma$. Finally, the last equation states that the multiplier $\Theta^I$, defined on $\Gamma$, must be equal to the jump in normal gradient of $\Theta$.

Turning now to the adjoint level set problem; due to the geometric non-linearity induced by $\phi$ on the domains of integration a careful treatment of each term is required as described in \cite{bernauer_optimal_2011}. Denoting $D \llbracket H(\phi); \delta \phi \rrbracket$ the variation of $H(\cdot)$ in the direction $\delta \phi$, the adjoint level set equations are derived by setting $\mathcal{L}_\phi(\cdot) = 0$
\begin{equation*}\label{eq:Lphi}
\arraycolsep=2pt\def\arraystretch{2.2}
\begin{array}{l}
\mathcal{L}_\phi \delta \phi = D \displaystyle \left\llbracket \mathcal{J}(T, \phi, w); \delta \phi \right\rrbracket \\
+ D \displaystyle \left\llbracket \int_0^{t_f} \int_{\Omega_S} \left( \frac{\partial T_S}{\partial t} - \nabla^2 T_S \right) \Theta_S; \delta \phi \right\rrbracket + D \displaystyle \left\llbracket \int_0^{t_f} \int_{\Omega_L} \left( \dfrac{\partial T_L}{\partial t} + \mathbf{u} \cdot \nabla T_L -\nabla^2 T_L \right) \Theta_L; \delta \phi \right\rrbracket \\
- D \displaystyle \left\llbracket \int_0^{t_f} \int_{\partial \Omega} \left( T_S - w \right) \Theta^D_S; \delta \phi \right\rrbracket - D \displaystyle \left\llbracket \int_0^{t_f} \int_{\partial \Omega} \left( T_L - T_b \right) \Theta^D_L; \delta \phi \right\rrbracket\\
- D \displaystyle \left\llbracket \int_0^{t_f} \int_\Gamma \left( T - T_M\right) \Theta^I; \delta \phi \right\rrbracket 
- D \displaystyle \left\llbracket \int_{0}^{t_{f}} \int_\Gamma \left(\dfrac{\partial \phi}{\partial t} + [\nabla T_i]^S_L \cdot \nabla \phi  \right) \psi; \delta \phi \right\rrbracket.
\end{array}
\end{equation*}
\change{We divide the contributions of the adjoint level set term-by-term. The contribution from the cost functional, removing the terms that do not depend on $\phi$, simplifies to
\begin{align*}
D \displaystyle \left\llbracket \mathcal{J}(T, \phi, w); \delta \phi \right\rrbracket  = D \displaystyle \left\llbracket \dfrac{\beta_2}{2} \int_{\Gamma^f} |\phi^f|^2; \delta \phi \right\rrbracket 
\end{align*} 
By using the theorem on the derivative of a boundary integral (Theorem \ref{theo:boundary_integral}), we obtain
\begin{align*}
D \displaystyle \left\llbracket \mathcal{J}(T, \phi, w); \delta \phi \right\rrbracket  = - \dfrac{\beta_2}{2} \int_{\Gamma^f} \frac{\delta \phi^f}{|\nabla \phi|} \left( \frac{\partial |\phi^f|^2}{\partial n} + \kappa |\phi^f|^2 \right).
\end{align*} 
By specifying that $\delta \phi = 0$ on $\delta \Omega$, the second and third terms are then equal to zero. For the fourth term, we need to assume that the adjoint level set $\phi$ and the multiplier $\Theta^I$ are defined on all of $\Omega$. Under such assumptions, we can apply the same boundary integral theorem (Theorem \ref{theo:boundary_integral}).  Moreover, using the fact that $T_M$ is constant along the interface, this term simplifies to
\begin{align*}
- D \displaystyle \left\llbracket \int_0^{t_f} \int_\Gamma \left( T - T_M\right) \Theta^I; \delta \phi \right\rrbracket = - \int_0^{t_f} \int_\Gamma \frac{\delta \phi}{|\nabla \phi|} \frac{\partial T}{\partial n} \Theta^I.
\end{align*} 
Finally, by applying the same boundary integral theorem together with the chain rule, the last term simplifies to
\begin{align*}
- D \displaystyle \left\llbracket \int_{0}^{t_{f}} \int_\Gamma \left(\dfrac{\partial \phi}{\partial t} + [\nabla T_i]^S_L \cdot \nabla \phi  \right) \psi; \delta \phi \right\rrbracket \\
= -\int_0^{t_f} \int_{\Gamma} \frac{\delta \phi}{|\nabla \phi|}\left(\frac{\partial}{\partial n}\left(\left( \frac{\partial \phi}{\partial t}+[\nabla T]^S_L \cdot \nabla \phi\right) \psi\right) + \kappa\left(\frac{\partial \phi}{\partial t}+[\nabla T]^S_L \cdot \nabla \phi\right) \psi \right)\\
+  \displaystyle \int_0^{t_f} \int_{\Gamma} \left( \frac{\partial \delta \phi}{\partial t}+[\nabla T]^S_L \cdot \nabla \delta \phi \right) \psi\\
= - \displaystyle \int_0^{t_f} \int_{\Gamma} \frac{\delta \phi}{|\nabla \phi|}\frac{\partial}{\partial n}\left( \frac{\partial \phi}{\partial t}+[\nabla T]^S_L \cdot \nabla \phi\right) \psi + \int_0^{t_f} \int_{\Gamma} \left( \frac{\partial \delta \phi}{\partial t}+[\nabla T]^S_L \cdot \nabla \delta \phi \right) \psi.
\end{align*} 
By using the assumption of constant velocity $v$ extended in the normal direction on all of $\Omega$, we obtain that
\begin{align*}
- D \displaystyle \left\llbracket \int_{0}^{t_{f}} \int_\Gamma \left(\dfrac{\partial \phi}{\partial t} + [\nabla T_i]^S_L \cdot \nabla \phi  \right) \psi; \delta \phi \right\rrbracket = - \displaystyle \int_0^{t_f} \int_{\Gamma} \frac{\delta \phi}{|\nabla \phi|}\left( \frac{\partial \delta \phi}{\partial t}+[\nabla T]^S_L \cdot \nabla \delta \phi\right) \psi.
\end{align*}
In order to obtain the adjoint Stefan condition, we need to move the derivatives from $\delta \phi$ to $\psi$. Assuming that $\psi$ is defined on all of $\Omega$, we can use the corollary \ref{coro:surface_transport} on integration by parts in time on a moving surface
\begin{align*}
- \displaystyle \int_0^{t_f} \int_{\Gamma} \frac{\delta \phi}{|\nabla \phi|}\left( \frac{\partial \phi}{\partial t}+[\nabla T]^S_L \cdot \nabla \phi\right) \psi = -\int_{\Gamma^f} \delta \phi^f \psi^f + \int_{\Gamma^0} \delta \phi^0 \psi^0\\
- \displaystyle \int_0^{t_f} \int_{\Gamma} \left( \delta \phi \frac{\partial \psi}{\partial t} + \nabla(\delta \phi \: \psi) \cdot v + \delta \phi \: \psi \operatorname{div}_\Gamma v + \psi [\nabla T]^S_L \cdot \nabla \delta \phi \right) \\
= \displaystyle -\int_{\Gamma^f} \delta \phi^f \psi^f + \int_{\Gamma^0} \delta \phi^0 \psi^0 - \int_0^{t_f} \int_{\Gamma} \delta \phi \frac{\partial \psi}{\partial t} \\
\displaystyle - \int_0^{t_f} \int_{\Gamma} \delta \phi \left(\nabla(\psi) \cdot v + \psi \operatorname{div}_\Gamma v \right) + \int_0^{t_f} \int_{\Gamma} \left( \psi [\nabla T]^S_L \cdot \nabla \delta \phi - \psi \nabla \delta \phi \cdot v \right).
\end{align*}
The last term cancels due to the fact that $[\nabla T]^S_L = v$ on $\Gamma$. The resulting terms of the adjoint level set, using $\delta \phi^0 = 0$, give
\begin{align*}
\displaystyle \mathcal{L}_\phi \delta \phi = - \dfrac{\beta_2}{2} \int_{\Gamma^f} \frac{\delta \phi^f}{|\nabla \phi|}  \frac{\partial |\phi^f|^2}{\partial n} + \kappa |\phi^f|^2\\
- \displaystyle \int_0^{t_f} \int_\Gamma \frac{\delta \phi}{|\nabla \phi|}  \frac{\partial T}{\partial n} \Theta^I \\
\displaystyle -\int_{\Gamma^f} \delta \phi^f \psi^f - \int_0^{t_f} \int_{\Gamma} \delta \phi \frac{\partial \psi}{\partial t} \\
\displaystyle - \int_0^{t_f} \int_{\Gamma} \delta \phi \left(\nabla(\psi) \cdot v + \psi \operatorname{div}_\Gamma v \right).
\end{align*}
By regrouping the terms by their domains of integration and requiring $\mathcal{L}_\phi \delta \phi = 0$ $\forall \: \delta \phi$, we obtain that
\begin{align*}
\psi^f = - \frac{1}{|\nabla \phi|} \dfrac{\beta_2}{2} \left( \frac{\partial |\phi^f|^2}{\partial n} + \kappa |\phi^f|^2 \right) \\
- \frac{\partial \psi }{\partial t} = \nabla \psi \cdot v + \psi \operatorname{div}_\Gamma v + \frac{1}{|\nabla \phi|} \frac{\partial T}{\partial n} \Theta^I.
\end{align*}
This equation can be interpreted as a first-order conservation law on $\Gamma$ and the source term on the right-hand-side can be extended in the neighborhood of $\Gamma$.}
For details on the derivation and simplifications of $\mathcal{L}_\phi(\cdot) = 0$, we refer the reader to~\cite{fullana2022}.

Finally, the last term to complete the adjoint Stefan problem is the gradient equation. By setting $\mathcal{L}_w \delta w = 0$ in Equation~\ref{eq:Lagrange}, we have
\begin{equation*}\label{eq:Grad}
\arraycolsep=2pt\def\arraystretch{2.2}
\begin{array}{l}
\displaystyle \mathcal{L}_w \delta w = \left\llbracket \frac{\beta_3}{2} \int_{0}^{t_{f}} \int_{\partial \Omega_S} |w|^2 ; \delta w \right\rrbracket - \left\llbracket \int_{0}^{t_{f}} \int_{\partial \Omega_S} \left( T_S - w \right) \Theta^D_S; \delta w \right\rrbracket.
\end{array}
\end{equation*}

Using the previously defined multiplier $\Theta_S^D$ that is identically equal to $\Theta_S$ on $\partial \Omega_S$ we obtain the gradient equation $0 = \beta_3 w + \Theta$ on $\partial \Omega_S$.
Putting it all together, we can now formulate the adjoint problem as follows 
\begin{center} 
Find a function $\Theta: \Omega \times[t_{f}, 0] \rightarrow \mathbb{R}$ and a function $\psi$ :
$\Omega \times[t_{f}, 0] \rightarrow \mathbb{R}$ such that
\[\arraycolsep=2pt\def\arraystretch{2.2}
\label{AP} \tag{AP} \left\{
\begin{array}{rcllr} - \dfrac{\partial \Theta_{S}}{\partial t} &=& \nabla^2 \Theta_{S} & \: \text{in} \: \Omega_S(t) & \text{(AP.a)} \\
- \dfrac{\partial \Theta_{L}}{\partial t} - \mathbf{u} \, \nabla \Theta_L &=& \nabla^2 \Theta_{L} & \: \text{in} \: \Omega_L(t) & \text{(AP.b)} \\ 
\Theta(x,t_f) &=& \beta_1 (T({t_{f}})-T_{t_{f}}) & \: \text{in}  \: \Omega & \text{(AP.c)} \\
\dfrac{\partial \Theta_S}{\partial n} &=& 0 & \: \text{on} \: \partial \Omega_S & \text{(AP.d)} \\
\dfrac{\partial \Theta_L}{\partial n} &=& 0 & \: \text{on} \: \partial \Omega_L & \text{(AP.e)} \\
\Theta(x,t) &=& \psi |\nabla \phi| & \: \text{on} \: \Gamma(t) & \text{(AP.f)} \\
\displaystyle \dfrac{\partial \psi}{\partial t} + \operatorname{div} (\psi v) &=& \dfrac{1}{|\nabla \phi |} \dfrac{\partial T}{\partial n} [\nabla \Theta^I]^S_L \cdot n & \: \text{on} \: \Gamma(t) & \text{(AP.g)} \\
\psi(x,t_f) &=& -\dfrac{\beta_2}{2 } \left( \dfrac{\partial }{\partial n}|\phi^f|^2 + \kappa |\phi^f|^2 \right)& \: \text{in} \: \Omega & \text{(AP.h)} \\
0 &=& \beta_4 w + \Theta & \: \text{on} \: \partial \Omega & \text{(AP.i)}
\end{array}
\right.
\]
\end{center}
We again emphasis on the fact that, here, the velocity field $\mathbf{u}$ is considered as a known variable, eliminating the need to derive the adjoint of the Navier-Stokes equations n the fluid phase and leading to an incomplete adjoint formulation.

\section{Optimization results}\label{sec:results}

\change{Adjoint-based methods—being efficient gradient-based optimization approaches—are often the methods of choice in PDE-constrained optimal control. By solving an adjoint equation once, one obtains the gradient of the objective with respect to all control variables at minimal cost, making the optimal-control formulation both tractable and efficient even for large-scale parameter spaces.} We choose to solve the minimization problem \eqref{MP} by using the limited memory BFGS (L-BFGS) method, a quasi-Newton method originally described in~\cite{Liu1989}. The main characteristic of this method is that it determines the descent direction by preconditioning the gradient with an approximation of the Hessian matrix. This information is obtained using past approximations -- the number of approximations is determined by the memory length parameter which is set to $m = 10$ -- as well as the gradient. As an initial guess for the initial Hessian, we use the scaled identity matrix as described in~\cite{Wright2006}. Algorithm~\ref{alg:LBFGS} summarizes the L-BFGS method used in our numerical example. The algorithm is stopped at a given iteration $n$ if one of the following criteria is fulfilled
\begin{itemize}
\item The relative difference in control variable $\left| \dfrac{w^{n-1} - w^{n}}{w^n} \right|  < 10^{-8}$.\\

\item The relative difference in cost functional $\left| \dfrac{\mathcal{J}^{n-1} - \mathcal{J}^{n}}{\mathcal{J}^n} \right|  < 10^{-8}$. This criterion can be relaxed to allow temporary increase of the cost functional, for example to ‘escape’ a local minimum.\\

\item The norm of the gradient $\left| \nabla \mathcal{J}^{n} \right|  < 10^{-6}$.
\end{itemize}
\begin{algorithm*}
    \SetKwData{Left}{left}
	\SetKwData{This}{this}
	\SetKwData{Up}{up}
	\SetKwInOut{Input}{input}\SetKwInOut{Output}{output}
	\Input{$w^0$, $m = 10$}
	\Output{$w$, $T$, $\phi$, $\Theta$, $\psi$}
	
	\BlankLine $k \leftarrow 0$, $l \leftarrow 0$

	\While{not converged}{
	\BlankLine Solve the forward Stefan problem (\ref{FP}) for $T^k$ and $\phi^k$ \\
	\BlankLine Solve the adjoint Stefan problem (\ref{AP}) for $\Theta^k$ and $\psi^k$\\
	\BlankLine Compute the gradient: $$\nabla \mathcal{J}^k = \beta_3 w^k + \Theta^k$$ \\
	\If{$k \geq 1$}{
	$s^{k-1} = \sigma^{k-1} d^{k-1} \quad g^{k-1} = \nabla \mathcal{J}^k - \nabla \mathcal{J}^{k-1}$ \\
	\uIf{$(s^{k-1})^{T} g^{k-1} \leq 0$}{$l \leftarrow 0$}
	\ElseIf{$(s^{k-1})^{T} g^{k-1} > 0$}{$l \rightarrow l + 1$\\
	\If{$l > m$}{Remove $\{ s^{l-m}, g^{l-m} \}$}
	Add $\{ s^{l-m}, s^{l-m}\}$\\
	}}
	\BlankLine Choose an initial approximation to the inverse of the Hessian $H^{k}_{0}$
	\BlankLine Construct the direction $d^k = -H^k \nabla \mathcal{J}^k$
	\BlankLine Determine $\sigma^k$ using a Line Search algorithm with backtracking where $\sigma^k = \operatorname{argmin}\mathcal{J}(w^k + \sigma^k d^{k})$
	\BlankLine Update $w^{k+1} = w^k + \sigma^k d^k$ 
	\BlankLine $k \rightarrow k + 1$
	}	
	\caption{Optimization procedure using the L-BFGS method}\label{alg:LBFGS}
\end{algorithm*}

To test our gradient-based minimization procedure, we now consider the numerical setup defined in~\ref{fig:schematic rayleigh}. The goal of this optimization procedure is to actuate on the top boundary condition to diminish the fluid layer growth, therefore reducing the onset of the instability. The actuator $w$ is parametrized with the following basis
\begin{equation}\label{eq:basis1}
w = -|a_1| - |a_2| \, (1 - \tanh(2\, x)^2), 
\end{equation}
where $x$ corresponds to the bounds of the domain and $a_1$ and $a_2$ to the basis coefficients. Through the optimization process, the amplitude of each coefficient is determined using the gradient equation~\eqref{AP}.i. \change{Practically, we fit the the temperature values on the boundary with a give basis, and update the coefficient at each iteration of the optimization procedure.} By opting for a parameterized distribution we ensure the smoothness of the actuation function. Indeed, due to the high sensitivity of the cost functional with respect to the basis considered -- too many parameters will create multiple local minima -- the number of parameters are kept at a low enough value to ensure the convexity of the problem while allowing spatial variation of the actuation function.

\begin{figure}[ht!]
\centering
\subfloat[Average height]{
\centering
\includegraphics[width=0.47\linewidth]{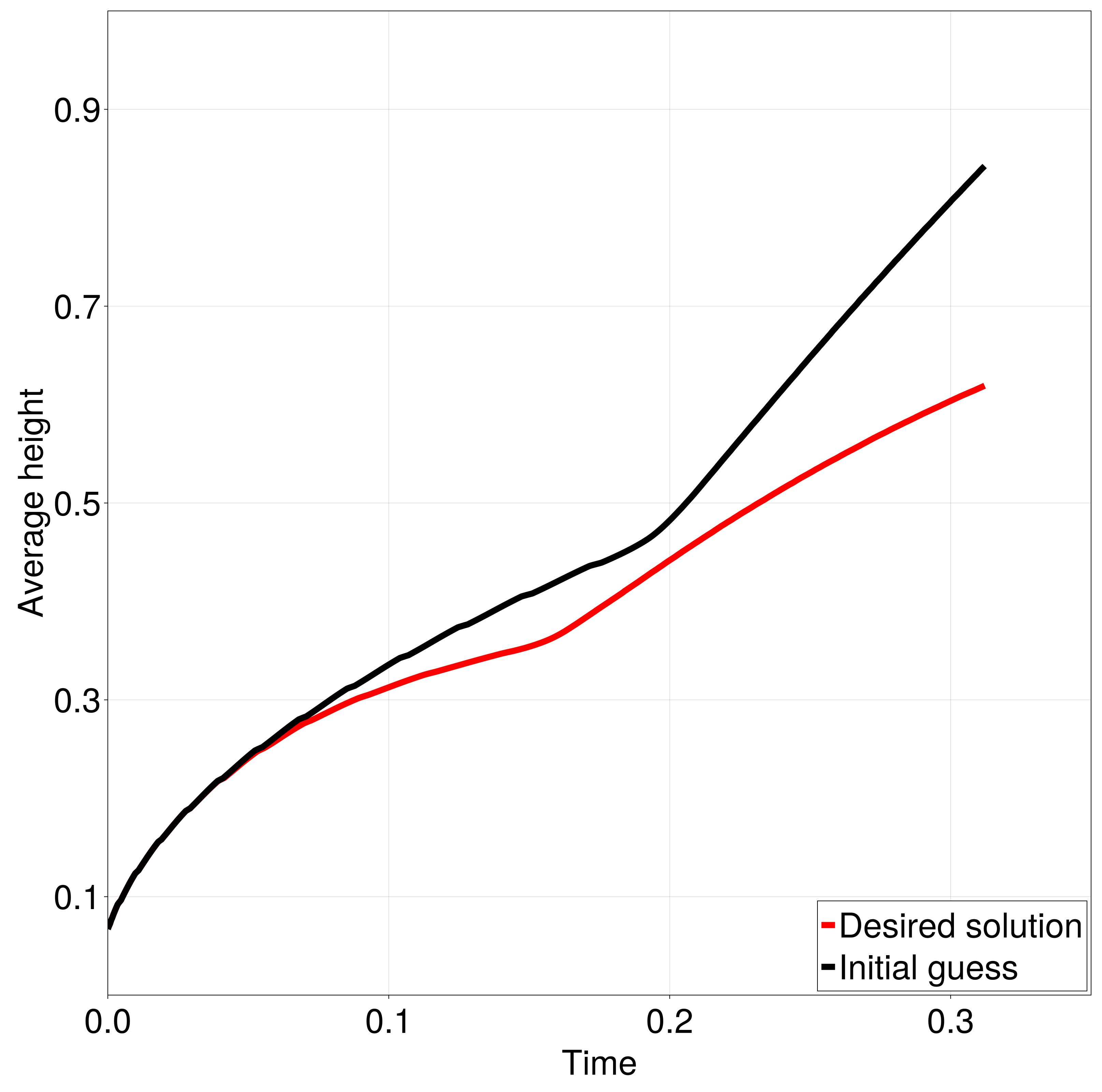}
}
\subfloat[Effective Rayleigh number]{
\centering
\includegraphics[width=0.47\linewidth]{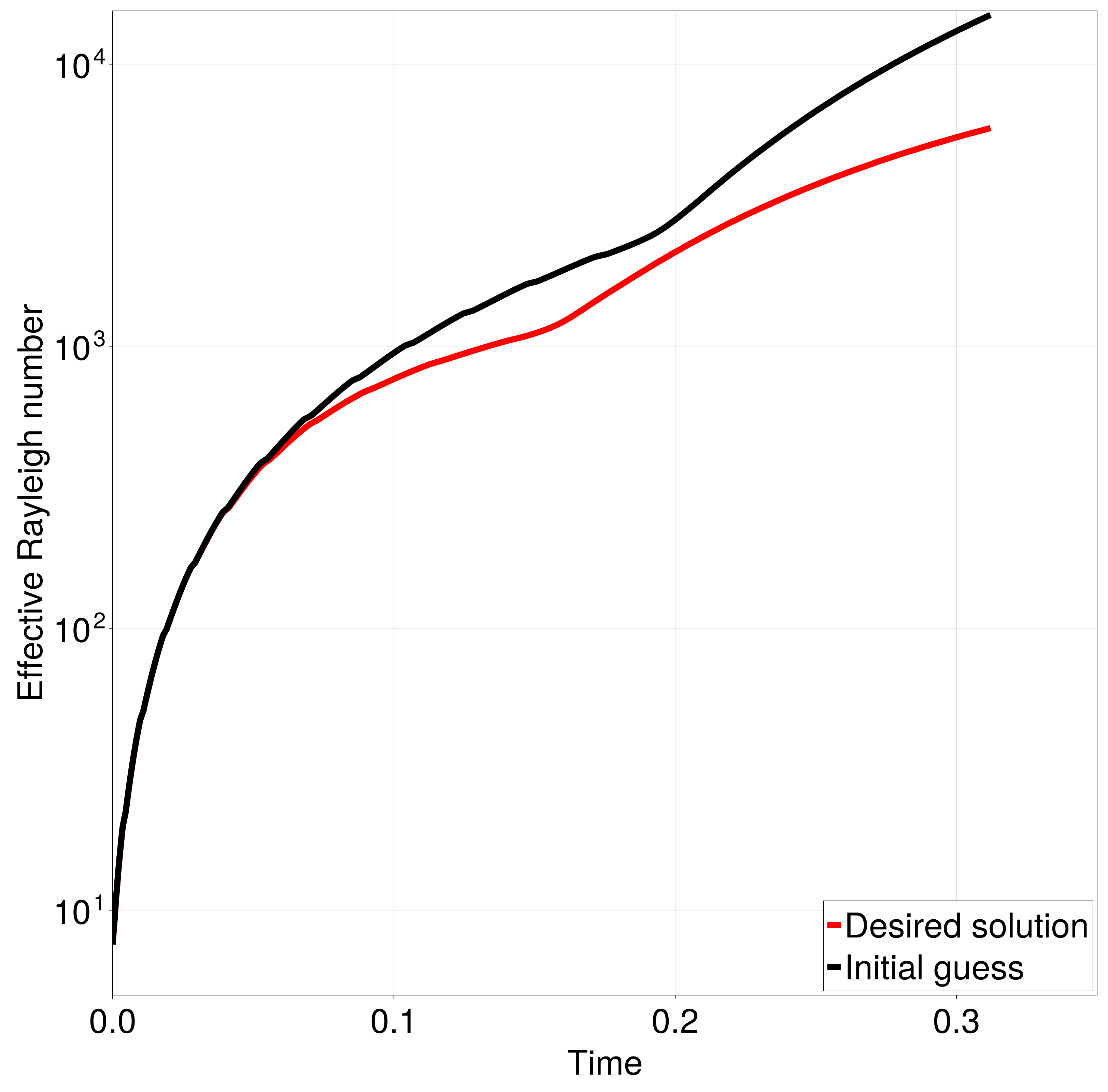}
}
    \caption{Average height and effective Rayleigh number as a function of time for the desired solution and initial guess.}
    \label{fig:hRa_adj}
\end{figure}

\begin{figure}[ht]
\centering
\subfloat[Iteration 0]{
\centering
\includegraphics[width=0.9\linewidth]{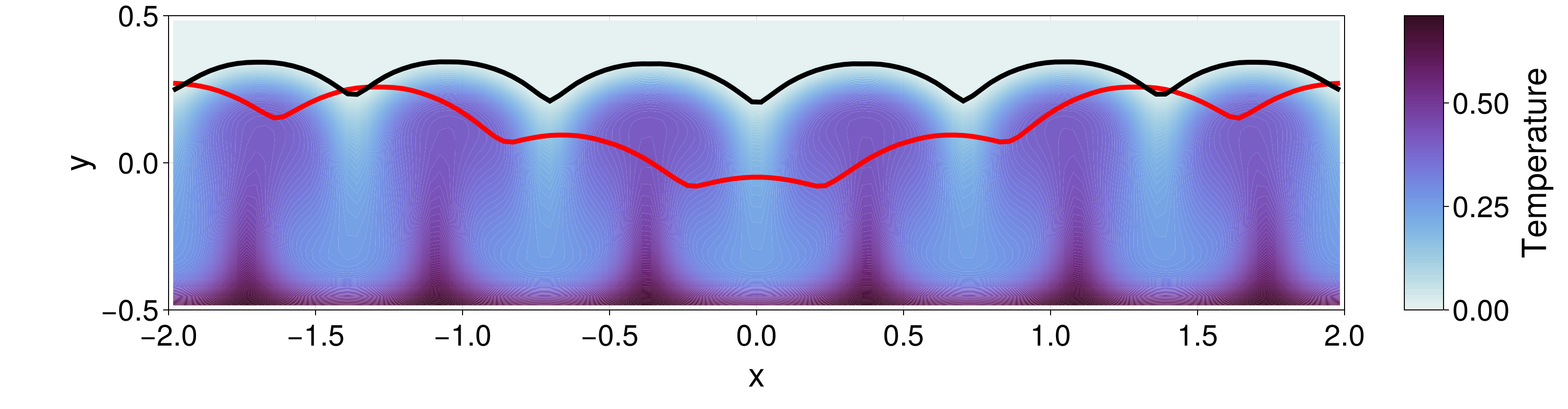}
}

\subfloat[Iteration 1]{
\centering
\includegraphics[width=0.9\linewidth]{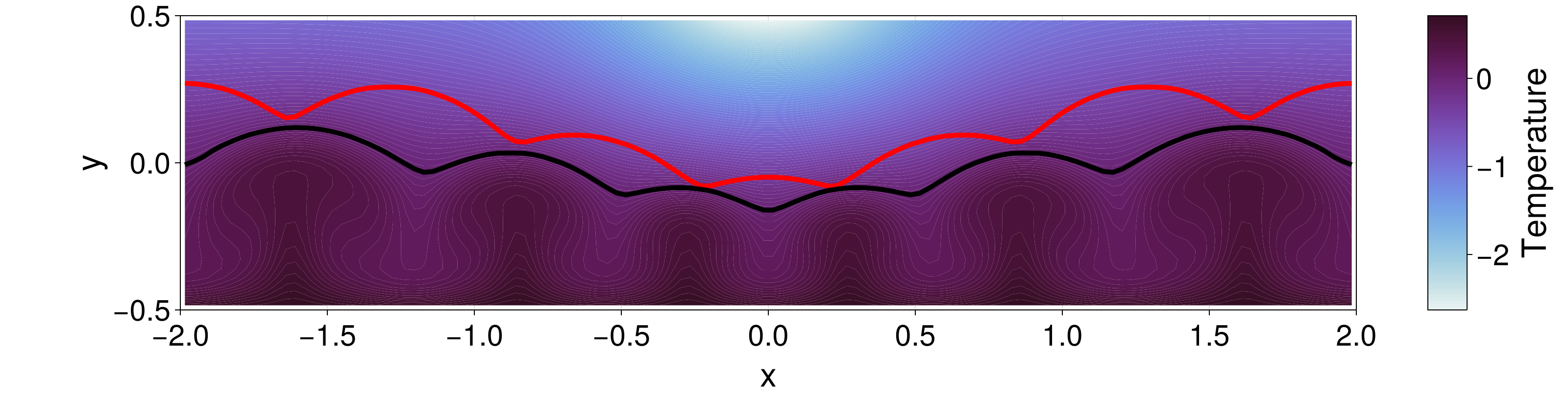}
}

\subfloat[Iteration 2]{
\centering
\includegraphics[width=0.9\linewidth]{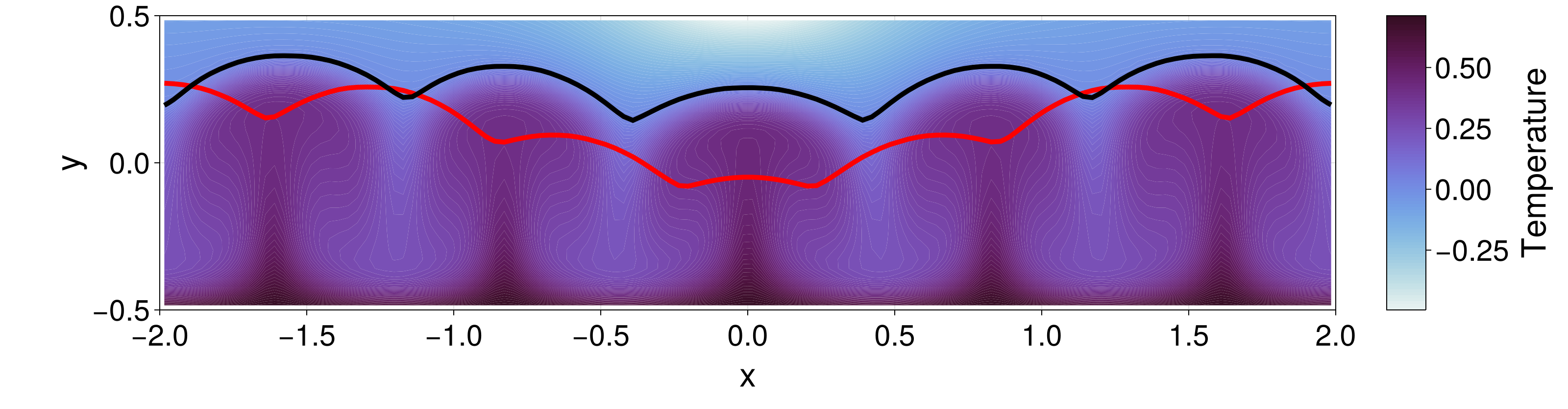}
}

\subfloat[Iteration 15]{
\centering
\includegraphics[width=0.9\linewidth]{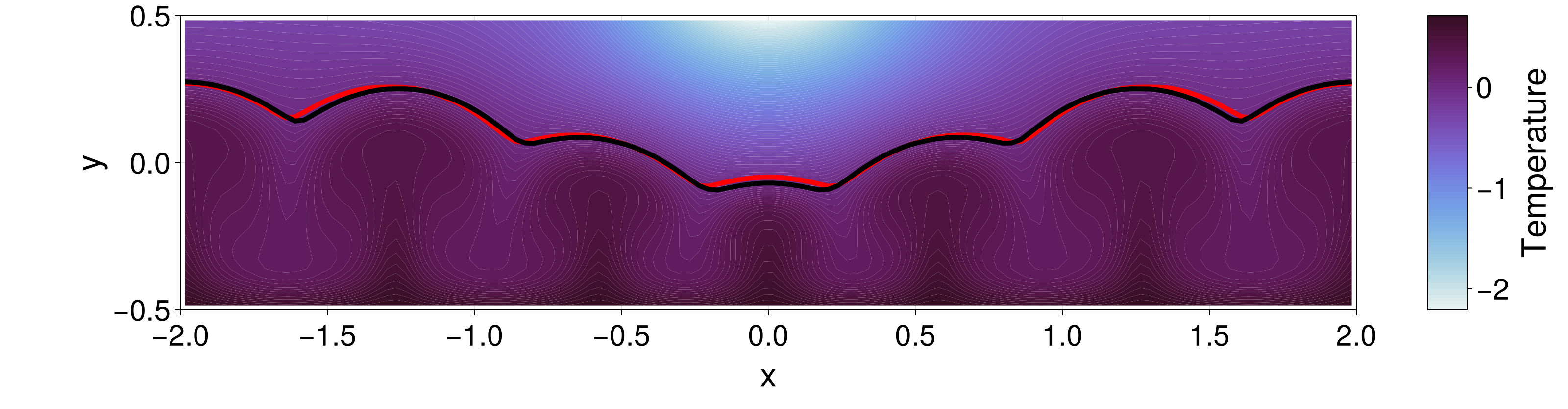}
}
    \caption{Iterations 0, 1, 2 and 15 of the optimization procedure. The red curve represents the desired shape and the black one the final position of the interface at a given iteration. The color map corresponds to the final temperature field of the \eqref{FP}.}
    \label{fig:RB}
\end{figure}

The global Rayleigh number is set to $\operatorname{Ra} = 10^5$ and an initial guess $w = 0$. The coefficients in the cost functional~\eqref{eq:cost_functional} are set to $\beta_1 = 1$, $\beta_2 = 1$ and $\beta_3 = 10^{- 3}$. The desired solution is computed beforehand with coefficients $a_1 = 0.3$ and $a_2 = 2$.
Figure~\ref{fig:RB} shows the interface position and temperature field at different iterations of the optimization procedure. The initial guess with $w = 0$ allows the interface to grow unperturbed, creating self-similar convection cells that affect the interface shape. The localized heat flux at the top boundary will create a heat gradient at the center of the domain leading to a destabilization of the interface and the appearance of the convection-driven regime before the unperturbed case (Fig.~\ref{fig:hRa_adj}).

\begin{figure}[ht!]
\centering
\includegraphics[width=0.8\textwidth]{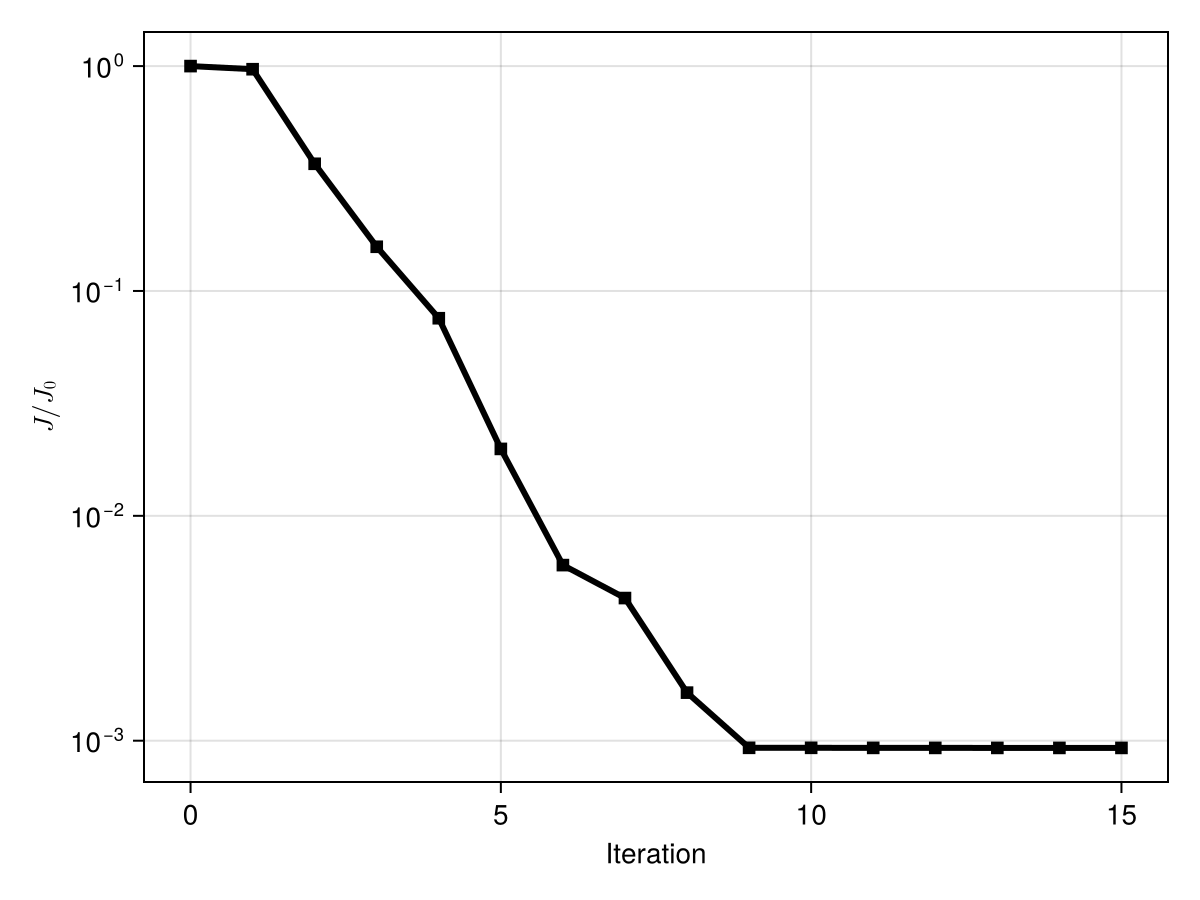}
\caption[Schematic of the melting boundary problem.]{Normalized cost functional as a function of the iterations of the L-BFGS optimization procedure \change{in the first test case}.}
\label{fig:cost}
\end{figure}

\begin{figure}[ht!]
\centering
\includegraphics[width=0.8\textwidth]{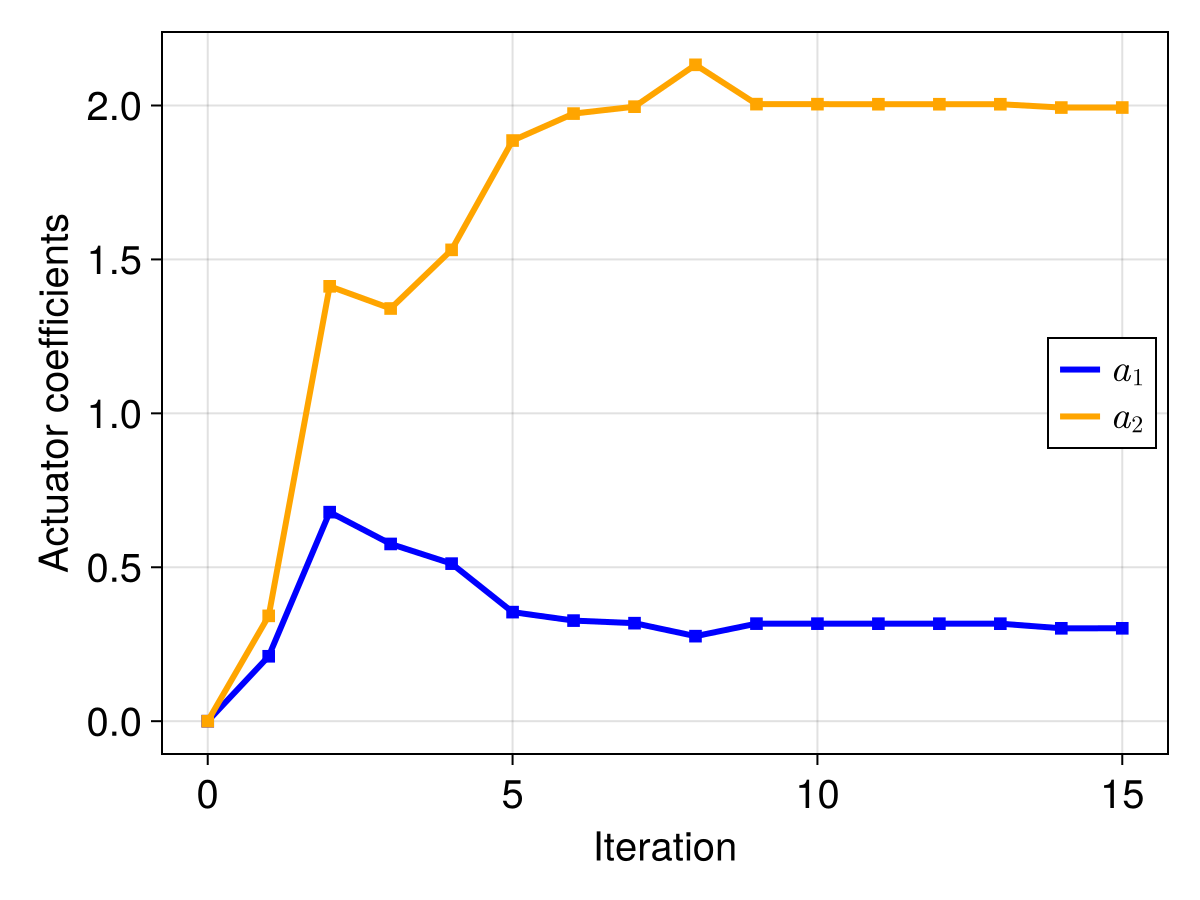}
\caption[Schematic of the melting boundary problem.]{\change{Basis coefficients as a function of the iterations of the L-BFGS optimization procedure in the first test case.}}
\label{fig:coeff}
\end{figure}

As the optimization procedure advances, we observe that the final interface position converges to the desired one, with a localized heat flux that prevents the melting interface growth at the center of the domain. In
Figure~\ref{fig:cost} we show the evolution of the normalized cost functional $\mathcal{J}/\mathcal{J}_0$ as a function of the iterations. The procedure stops at iteration 15, with a relative value three orders of magnitude lower, as the norm of the gradient goes below the prescribed threshold \change{and the basis coefficients have reached the desired ones (Fig.\ref{fig:coeff}).}

\begin{figure}[ht!]
\centering
\includegraphics[width=0.8\textwidth]{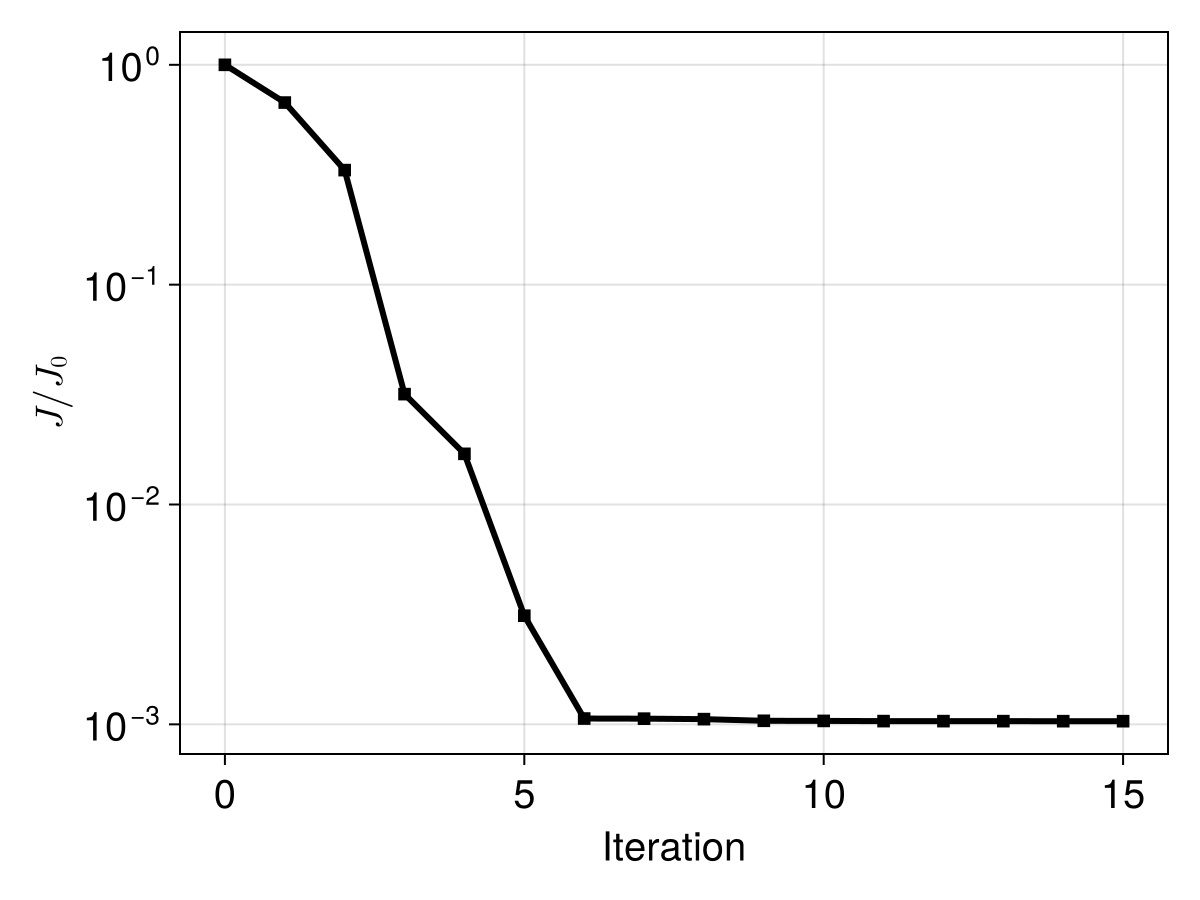}
\caption[Schematic of the melting boundary problem.]{\change{Normalized cost functional as a function of the iterations of the L-BFGS optimization procedure in the second test case}.}
\label{fig:cost2}
\end{figure}

\begin{figure}[ht!]
\centering
\includegraphics[width=0.8\textwidth]{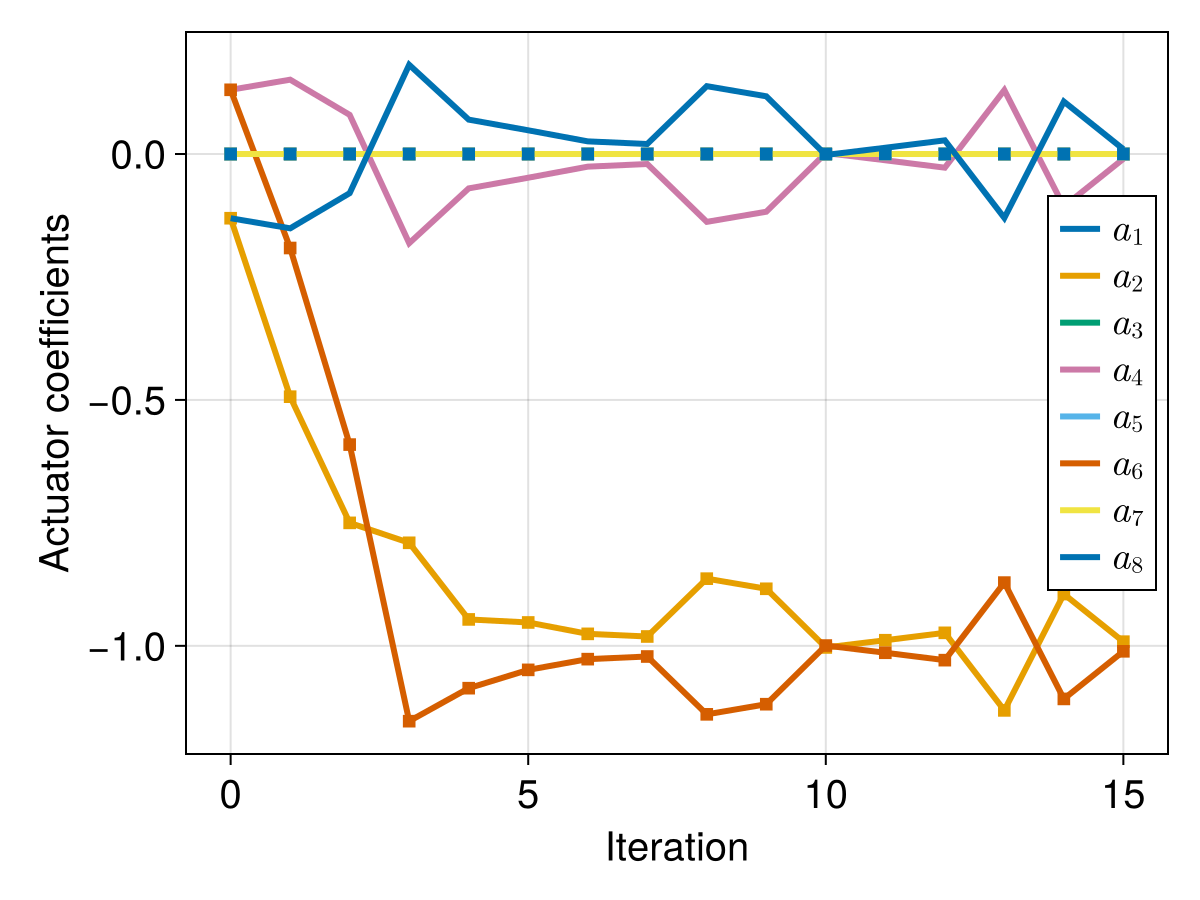}
\caption[Schematic of the melting boundary problem.]{\change{Basis coefficients as a function of the iterations of the L-BFGS optimization procedure in the second test case.}}
\label{fig:coeff2}
\end{figure}

\change{We now show a second optimization example, with a more complex basis
\begin{equation}\label{eq:basis2}
w = \sum_{n=1}^{4} a_{n}\,\sin^{n}(2\pi x)
+ \sum_{n=1}^{4} a_{n + 4}\,\cos^{n}(2\pi x). 
\end{equation}
The global Rayleigh number is identical to the previous test case. The initial guess is 0 and the desired solution is such that the interface remains flat, with a constant heat flux $w = -1$. As shown in Figs.\ref{fig:cost2} and \ref{fig:coeff2}, the optimization procedure is a able to recover the desired solution even with a higher number of fitting coefficients.}

\change{We now compare the results to a derivative-free method (where no information on the gradient is required),  the Particle Swarm optimization algorithm~\cite{ParticleSwarm}. This method attempts to improve global convergence by switching between four evolutionary states: exploration, exploitation, convergence, and jumping out. In the jumping out state it tries to take the best particle and move it away from its local optimum, to improve the ability to find a global one. 
We test the derivative-free optimization procedure on the cases presented using the same parameters for the cost functional. Table~\ref{tab:opt2} summarizes the comparison between both methods. As expected, the derivative-free one attains a lower minimum at a cost of a much higher number of function evaluations (one order of magnitude higher). \changee{The L-BFGS algorithm driven by the incomplete adjoint thus attains objective values close to those of the particle-swarm optimizer at a fraction of the computational cost.}}
\begin{table*}[hbt]
\setlength{\tabcolsep}{0.5pc}

\begin{tabular*}{\textwidth}{@{}l@{\extracolsep{\fill}}ccccccc}
\hline
                 \multicolumn{1}{c}{Case} 
                 & \multicolumn{3}{c}{L-BFGS} 
                 & \multicolumn{2}{c}{Particle Swarm}\\
                 & \multicolumn{1}{c}{$\mathcal{J}$ calls}
                 & \multicolumn{1}{c}{$\nabla \mathcal{J}$ calls} 
                 & \multicolumn{1}{c}{$\mathcal{J}_{final}/\mathcal{J}_{0}$}
                 & \multicolumn{1}{c}{$\mathcal{J}$ calls} 
                 & \multicolumn{1}{c}{$\mathcal{J}_{final}/\mathcal{J}_{0}$}\\
\hline
Case 1  & $ 31$ & $16$  & $  1.02 \times 10^{-3}$ & $928$ & $2.04 \times 10^{-4}$ \\
Case 2 & $ 42$ &  $  16$ & $ 1.01 \times 10^{-3}$ & $2321$ & $2.14 \times 10^{-4}$ \\
\hline
\end{tabular*}
\caption{Comparison between the L-BFGS and the Particle swarm method for the two considered cases.}
\label{tab:opt2}
\end{table*}

\FloatBarrier
\section{Conclusion}

This paper demonstrates the feasibility of extending the optimization of Stefan problems to scenarios where the flow in the fluid phase is significant and cannot be ignored. By treating the velocity field as a known variable in the adjoint equations, we were able to derive an incomplete continuous adjoint formulation for the two-phase problem. Our results indicate that the adjoint-based gradient descent method effectively controls the shape of melting fronts in this highly nonlinear system, \changee{while greatly reducing the number of function evaluations compared with a derivative-free particle-swarm algorithm. The present formulation is limited to two-dimensional, laminar flows with a prescribed top boundary temperature in a periodic domain and an incomplete adjoint in which the Navier–Stokes adjoint problem is not solved. While the numerical examples indicate that the resulting gradients are accurate enough for practical optimization, the method does not capture sensitivities associated with changes in the flow field itself. This approach nevertheless shows great potential for applications in more complex and dynamic melting processes. Future work could explore more complex actuator basis, non-periodic domains, and three-dimensional systems to further enhance the optimization framework and broaden its applicability.}

\subsection*{Funding}
There is no source of funding.

\subsection*{Author contributions}
T.F. and T.S. conceived the study. T.F. performed the numerical simulations. T.F. wrote the original draft with feedback from all authors. T.S. supervised the study.

\subsection*{Data availability}
The data used in this study are available from the corresponding authors upon request.

\subsection*{Replication of results}
The results are obtained with an in-house Julia code Flower.jl \href{https://github.com/flnt/Flower.jl/}{https://github.com/flnt/Flower.jl/}.

\subsection*{Ethics approval and Consent to participate}
Not applicable.

\subsection*{Conflict of interest}
The authors declare no competing interests.

\appendix
\section{\change{Shape calculus theorems}}

\change{\begin{theorem}[Derivative of boundary integral]\label{theo:boundary_integral}
Let $J(\Omega) = \int_\Gamma f ds$ be a boundary integral, the derivative is given by
$$
\begin{array}{ccc}
d J(\Omega ; V)&=&\left.\left(\dfrac{d}{d \lambda} \displaystyle\int_{\Gamma_{\lambda}} f d s_{\lambda}\right)\right|_{\lambda=0}\\
&=&-\displaystyle\int_D \dfrac{\delta \phi}{|\nabla \phi|}\left(\dfrac{\partial f}{\partial n}+f \kappa\right) ds
\end{array}
$$
where $V$ is the velocity field and $\kappa$ is the mean curvature of $\Gamma$.
\end{theorem}
\begin{theorem}[Surface transport theorem]\label{theo:surface_transport}
Let $f(\cdot, t): \mathcal{S}_{t} \rightarrow \mathbb{R}$ be a scalar field defined on the moving surface $\mathcal{S}_{t} .$ Then
$$
\frac{d}{d t} \int_{\mathcal{S}_{t}} f(x, t) d \mathcal{S}_{t}=\int_{\mathcal{S}_{t}} \dot{f}(x, t)+f(x, t) \operatorname{div} \mathcal{S}_{t} \mathbf{w}(x, t) d \mathcal{S}_{t}
$$
where $\mathbf{w}$ is the normal velocity of the moving surface $\mathcal{S}_{t}$ and $\dot{f}$ is the parameter-time derivative of $f$. If $f(\cdot, t)$ is the restriction of a function $\hat{f}(\cdot, t)$ to $\mathcal{S}_{t}$, then
$$
\dot{f}(x, t)=\mathbf{w}(x, t) \cdot \nabla \hat{f}(x, t)+\frac{\partial}{\partial t} \hat{f}(x, t)
$$
\end{theorem}
\begin{lemma}[Integration by Parts in Time on a Moving Surface]\label{coro:surface_transport} 
$$
\begin{array}{c}
\displaystyle\int_{0}^{T} \int_{\mathcal{S}_{t}} \hat{g}(x, t) \hat{h}_{t}(x, t) d \mathcal{S}_{t} d t=\int_{\mathcal{S}_{T}} g(x, T) h(x, T) d \mathcal{S}_{T}-\int_{\mathcal{S}_{0}} g(x, 0) h(x, 0) d \mathcal{S}_{0} \\
-\displaystyle\int_{0}^{T} \int_{\mathcal{S}_{t}} \hat{g}_{t}(x, t) \hat{h}(x, t)+\mathbf{w}(x, t) \cdot \nabla(\hat{g}(x, t) \hat{h}(x, t)) \\
+\displaystyle g(x, t) h(x, t) \operatorname{div}_{\mathcal{S}_{t}} \mathbf{w} d \mathcal{S}_{t} d t
\end{array}
$$
where $g$ and $h$ are restrictions of $\hat{g}$ and $\hat{h}$ to $\mathcal{S}_{t}$.
\end{lemma}}

\FloatBarrier
\bibliography{biblio}

\end{document}